\documentclass[onecolumn]{IEEEtran}

\usepackage{amsmath,amssymb}
\usepackage{amsfonts}
\usepackage{mathrsfs}
\usepackage{extarrows}
\usepackage{graphicx}

\newtheorem{thm}{Theorem}[section]

\newtheorem{lem}{Lemma}[section]
\newtheorem{rem}{Remark}[section]

\newtheorem{assu}{Assumption}[section]

\def\R{\mathbb{R}}
\def\v{\varepsilon}
\def\t{\theta}
\def\g{\gamma}

\def\l{\lambda}

\IEEEoverridecommandlockouts

\title{Data Rate for Distributed Consensus of Multi-agent Systems with High Order Oscillator Dynamics}

\author{Zhirong Qiu$^{1}$, Lihua Xie$^{\dag,1}$, and Yiguang Hong$^{2}$ 
\thanks{*This work is partially supported by National Research Foundation of
Singapore under grant NRF-CRP8-2011-03 and NNSF of China under Grant
61174071. The work in this paper
has been partially presented in \cite{qiu_data_2014}.}
\thanks{$^{\dag}$Author for correspondence.}
\thanks{$^{1}$Z. Qiu and L. Xie are with School of Electrical and Electronic Engineering, Nanyang
Technological University, Singapore 639798.
     {\tt\small Email: qiuzhr@gmail.com; elhxie@ntu.edu.sg}}%
\thanks{$^{2}$Y. Hong is with the Key Laboratory of Systems and Control, Institute of Systems Science,
Chinese Academy of Sciences, Beijing 100190, China.
        {\tt\small Email: yghong@iss.ac.cn}}%
}

\begin{document}
\let\displaystyle\textstyle
\maketitle \thispagestyle{empty} \pagestyle{empty}

\begin{abstract}
Distributed consensus with data rate constraint is an important
research topic of multi-agent systems. 
Some results have been obtained for
consensus of multi-agent systems with integrator dynamics, but it
remains challenging for general high-order systems, especially in the presence of unmeasurable states. 
In this paper, we study the quantized consensus problem for a special
kind of high-order systems and investigate the corresponding data rate required
for achieving consensus. The state matrix of each agent is a $2m$-th
order real Jordan block admitting $m$ identical pairs of conjugate
poles on the unit circle; each agent has a single input, and only
the first state variable can be measured. The case of harmonic
oscillators corresponding to $m=1$ is first investigated under a
directed communication topology which contains a spanning tree,
while the general case of $m\ge 2$ is considered for a connected and
undirected network. In both cases it is concluded that the
sufficient number of communication bits to guarantee the consensus
at an exponential convergence rate is an integer between $m$ and
$2m$, depending on the location of the poles.
\end{abstract}

\section{Introduction}
Distributed consensus is a basic problem in distributed control of multi-agent systems, 
which aims to reach an interested common value of the states for a team of agents or
subsystems by exchanging information with their neighbors. 
A variety of consensus protocols have been proposed for different kinds of applications;
see the survey papers \cite{olfati-saber_consensus_2007,cao_overview_2013,knorn_overview:_2015}
and the reference therein.
Nonetheless, to apply the consensus protocol in a digital network with limited bandwidth,
it is necessary to introduce quantization and devise the corresponding encoding-decoding scheme. 
With static uniform quantization, quantized consensus was first studied in \cite{kashyap_quantized_2007} to achieve the approximate average consensus for integer-valued agents by applying gossip algorithms.
For a large class of averaging algorithms of real-valued agents, 
\cite{nedic_distributed_2009} established the bounds of the steady-state error
and the convergence times, as well as their dependence on the number of quantization levels.
Logarithmic quantizers with infinite quantization levels were adopted in \cite{carli_quantized_2010}
to guarantee the asymptotic average consensus.
To achieve the asymptotic average consensus with finite quantization levels,
a static finite-level uniform quantizer with a dynamic encoding scheme was proposed in \cite{li_distributed_2011-1},
and used to shown that an exponentially fast consensus
can be ensured by finite-level quantizers for multi-agent systems with general linear dynamics, 
whether the state is fully measurable \cite{keyou_you_network_2011},
or the state is only partially measurable and yet detectable \cite{meng_coordination_2015}. 
However, the lower bound of sufficient data rate for the consensus obtained in these works 
are overly conservative, and it is more appealing to achieve the consensus with fewer bits 
of information exchange from the perspective of reducing communication load.

Some works have been devoted to exploring the sufficient data rate to guarantee the consensus
of multi-agent systems with integrator dynamics, 
and single-integrator systems receive the most attention.
With a presumed bound of the initial state of each agent, Li et. al.
\cite{li_distributed_2011-1} showed that the average consensus can be achieved by 1
bit of information exchange for a fixed and undirected network,
which was further extended to the case when the network is balanced and contains a spanning tree \cite{zhang_quantized_2013}.
In an undirected network where the duration of link failure is bounded,
5-level quantizers suffices for the consensus \cite{li_distributed_2011-1}, 
which also holds when the network is periodically strongly connected \cite{li_quantized_2014}.
With a novel update protocol carefully screening the quantized message,
the presumed bound of initial values was shown to be unnecessary in \cite{olshevsky_consensus_2014}
and it was concluded that ternary messages
are sufficient for the average consensus under a periodically connected network.
Then, for double-integrator systems with only position being measurable,
\cite{tao_li_distributed_2012} concluded that 2 bits of communications suffice for the
consensus. By employing a totally different technique based on
matrix perturbation, $n$ bits were found to be sufficient to
achieve the consensus of multi-agent systems with $n$-th integrator
agent dynamics in \cite{qiu_quantized_2016}.
Still, it is unclear about the sufficient data rate to guarantee
the consensus for general high-order systems, especially when the
state variables are only partially measured. 


In this paper, we explore the data rate problem in achieving
quantized consensus of another kind of discrete-time high-order critical systems
as a complement of integrator systems.
The dynamics of each agent is described by a
$2m$-th order real Jordan block admitting $m$ identical pairs of conjugate
poles on the unit circle with single input, and only the first state
variable can be measured. We design the encoding-decoding scheme on the basis of
the constructability of the state variables of each
individual system: at each time instant
the quantizer will produce a signal to make an estimate for the current
measurable state, which is combined with the previous $ 2m-1 $ estimates of the measurable state to
obtain the estimate of the current full state. 
The same quantized signal will also be sent to neighbor
agents to generate the identical estimate of state. The control input 
is constructed in terms of the estimate of its own state, as well as
those of its neighbor agents. For harmonic oscillators ($ m=1 $), it is shown
that 2 bits of communications suffice to guarantee the exponentially fast consensus for
a directed network containing a spanning tree. For higher-order case
of $m\ge 2$, the exponentially fast consensus can be achieved with at most $2m$ bits under
an undirected network, provided that the undirected communication
topology is connected. The exact number of bits for achieving
consensus in both the cases is an integer between $m$ and $2m$,
depending on the frequency of oscillators or the location of 
poles on the unit circle.

Although the analysis of consensus and data rate in this paper
employs similar perturbation techniques as in \cite{qiu_quantized_2016}, the
problem posed here is much different, and it is much more
challenging to obtain an explicit data rate required for consensus
in the oscillator case (corresponding to complex eigenvalues).
In contrast to \cite{qiu_quantized_2016} where the special structure of integrator
dynamics enables a direct connection between the encoder's past
outputs and those at the present moment which leads to a convenient
iteration in the encoding scheme, a similar iteration is no longer
available for the estimation of state variables in the case of
oscillator dynamics. As
such, a new observer-based encoding scheme is devised. However, such an
encoding scheme leads to the involvement of control inputs into the estimation error, 
which makes the consensus analysis challenging. 
Furthermore, the expression of data rate for the oscillator case
requires calculating a linear combination of some rows of a matrix
which is a multiplication of the $(2m-1)$-th power of the system
matrix and the inverse of the observability matrix, and is hard to
obtain by a direct computation. To overcome this difficulty, we
transform the linear combination into a set of linear equations and
employ techniques of combinatorics. It is shown that a data rate
between $m$ and $2m$, depending on the frequencies of the
oscillations, suffices to achieve the consensus.
It is worthy noting that the result not only provides a sufficient
data rate for consensus of the systems under consideration but also
reveals an interesting connection between the data rate and the
system dynamics. We believe it will shed some further light on the data rate
problem for multi-agent systems of general dynamics.

The rest of the paper is organized as follows. Some preliminaries
about graph theory and the problem formulation are presented
in Section \ref{sec:problem}. Then the data rate problem for distributed consensus
of the coupled harmonic oscillators is conducted in Section \ref{sec:2order}, which
is followed by the general case of $m\ge 2$ in Section \ref{sec:high order}. 
For illustration, a numeric example is given in Section \ref{sec:example}.
Some concluding remarks are drawn in Section \ref{sec:conclusion}. The proofs of the
main lemmas can be found in the Appendix. 

Some notations listed below will be used throughout this paper. For a
matrix $U$, $U(i,j)$ and $U(i,\cdot)$ respectively denote its
$(i,j)$-th entry and $i$-th row; $U^T$ is its transpose, and $||U||$
is its infinity-norm. $\mathbb{N}^+$ is the set of positive
integers, and $\lceil a \rceil$, $\lfloor a\rfloor$ respectively
denote the smallest integer not less than $a$, and the largest
integer not greater than $a$. $C(n,k)$ is the number of
$k$-combinations from a given set of $n$ elements. $\textbf{1}_N$ is
the $N$ dimensional vector with every component being 1, and $I_m$ is
the identity matrix of order $m$. $\jmath=\sqrt{-1}$ is the unit
imaginary number. $J_{\lambda,n}$ denotes the $n$ dimensional Jordan
block with eigenvalue $\lambda$. $A\otimes B$ denotes the Kronecker
product between matrices $A$ and $B$. $\langle\cdot,\cdot\rangle$
denotes the standard inner product in Euclidean spaces.

\section{Problem Formulation}\label{sec:problem}
Consider a multi-agent system in the following form:
\begin{equation}\label{conjudynamics}
\left\{\begin{aligned}
x^{i}(t+1)&=Ax^{i}(t)+bu_{i}(t),\\
y_i(t)&=x_{i1}(t),
\end{aligned}\right.
\end{equation}
where $x^{i}(t)=\left[\,
                    x_{i1}(t), x_{i2}(t), \dots, x_{i,2m}(t)
                \right]^T\in \R^{2m}$,
$y_i(t), u_i(t)\in \R,\, i=1,\dots,N$ represent the state, output
and input of agent $i$, respectively. Moreover, $A=\left[
           \begin{array}{cccc}
             Q & I_2 & &  \\
               & Q & \ddots & \\
               &   & \ddots & I_2 \\
               &   & & Q\\
           \end{array}
         \right]\in \R^{2m\times 2m}$
 is a real Jordan form consisting of
$m$ pairs of conjugate eigenvalues $\cos\t+\jmath\sin\t$ with
$\sin\t\neq 0$ and $Q=\left[
          \begin{array}{cc}
            \cos \t & \sin \t \\
            -\sin \t & \cos \t \\
          \end{array}
        \right]$; $b=\left[\,
                    0,  \dots, 0, 1
                \right]^T\in \R^{2m}$.

Suppose that the total number of agents is $N$. Assumed to be
error-free, the digital communication channels between agents are
modeled as edges of a directed or undirected graph. A graph
$\mathcal{G}$ consists of a node set $\mathcal{V}=\{1,\dots,N\}$ and
an edge set $\mathcal{E}=\{(i,v): i,v\in \mathcal{V}\}$ where
self-loop $(i,i)$ is excluded. An edge $(i,v)$ of a directed graph
implies that node $v$ can receive information from node $i$, but not
necessarily vice versa. In contrast, for an undirected graph,
$(i,v)\in \mathcal{E}$ means mutual communications between $i$ and
$v$. For node $i$, $\mathcal{N}_{i}^{+}=\{v: (v,i)\in \mathcal{E}\}$
and $\mathcal{N}_{i}^{-}=\{v: (i,v)\in \mathcal{E}\}$ respectively
denote its in-neighbors and out-neighbors, which coincide if
$\mathcal{G}$ is undirected, and will be denoted as
$\mathcal{N}_{i}$. A directed path $(i_1,i_2),(i_2,i_3),\dots$ is
formed by a sequence of edges. For a directed graph $\mathcal{G}$,
if there exists a directed path connecting all the nodes, then
$\mathcal{G}$ is said to contain a spanning tree, which is
equivalent to the case of being connected when $\mathcal{G}$ is undirected.

Usually, a nonnegetive matrix $G=[g_{iv}]\in \R^{N\times N}$ is
assigned to the weighted graph $\mathcal{G}$, where $g_{iv}>0$ if
and only if $(v,i)\in \mathcal{E}$, and $g_{iv}=g_{vi}$ is further
required for an undirected graph. The connectivity of $\mathcal{G}$
can be examined from an algebraic point of view, by introducing the
Laplacian matrix $L=D^{G}-G$, where
$D^{G}=\text{diag}(d_{1}^{G},\dots,d_{N}^{G})$ and
$d_{i}^{G}=\sum_{v=1}^{N}g_{iv}$. By $L\mathbf{1}_N=0$, $L$ has at
least one zero eigenvalue, with the other non-zero eigenvalues on
the right half plane. $L$ has only one zero eigenvalue
if and only if $\mathcal{G}$ contains a spanning tree \cite{wei_ren_consensus_2005}.
We can always find a nonsingular matrix
$U_L=[\phi_1~\phi_2~\dots~\phi_N]$ with $\phi_1=\mathbf{1}_N/\sqrt
N$ and $||\phi_i||=1$, such that
$U_{L}^{-1}LU_{L}=\text{diag}\{J_{0,N_1},J_{\lambda_2,N_2},\dots,J_{\lambda_l,N_l}\}\triangleq
L_J$, where $0\le\text{Re}\l_2\le\dots\le\text{Re}\l_l$ with $\l_i$
being an eigenvalue of $L$. In particular, we denote
$\psi_i=U_{L}^{-1}(i,\cdot)^T$. Moreover,
$L_J=\text{diag}\{0,\l_2,\dots,\l_N\}$ with $0\le
\lambda_2\le\dots\le \l_N$ and $\psi_i=\phi_{i}^{T}$ if
$\mathcal{G}$ is undirected.


We adopt the following finite-level uniform quantizer $q_t (\cdot)$ in the
encoding scheme, where $M(t)\in \mathbb{N}^+$:
\begin{equation}\label{quantizer}
q_t(y)=\left\{\begin{array}{ll}
           0, & -\frac{1}{2}<y<\frac{1}{2}; \\
           j, & \frac{2j-1}{2}\leq y< \frac{2j+1}{2},~j=1,\dots,M(t)-1; \\
           M(t), & y\geq \frac{2M(t)-1}{2}; \\
           -q_t(-y),&y\leq -\frac{1}{2}.
         \end{array}\right.
\end{equation}

\begin{rem}\label{bits}
Clearly, the total number of quantization levels of $q_t (\cdot)$ is
$2M(t)+1$. Demanding that agent $i$ does not send out any signal
when the output is zero, it is enough to use
$\lceil\log_2(2M(t))\rceil$ bits to represent all the signals.
\end{rem}

The problem of distributed quantized consensus is solved if we can
design a distributed control protocol based on the outputs of the
encoding-decoding scheme, making the states of different agents
reach the agreement asymptotically:
\begin{equation}\label{leaderless}
 \lim_{t\to \infty}  [x^i(t)- x^j(t)]=0,\quad
i,j=1,2,\dots,N.
\end{equation}

\section{Harmonic Oscillator Case}\label{sec:2order}

In this section, we will start with the harmonic oscillator case as an example
to investigate how many bits of information exchange are enough to achieve consensus
exponentially fast with quantized neighbor-based control.
We separate it from higher-order cases due to its speciality and simplicity:
the solution of this basic case not only provides a result under a directed communication topology,
but also serves to facilitate the understanding of higher-order cases.
Some relevant remarks will be included in the next section, as a comparison between second-order
and higher-order cases, or a summary of general cases.
Note that now the system matrix $A=\left[
          \begin{array}{cc}
            \cos \t & \sin \t \\
            -\sin \t & \cos \t \\
          \end{array}
        \right]$.

\subsection{Encoding-decoding scheme and distributed control law}

An encoding-decoding scheme has a paramount importance in the quantized consensus, 
which should not only provide estimates
for all the states from the partially measurable states, but also
help reduce the data rate. Accordingly,
the encoder should serve as an observer based on iterations. To be
specific, inspired by the constructability in the sense that the
present state of the system can be recovered from the present and
past outputs and inputs, namely
\begin{equation}\label{observer}
\begin{aligned}
    \left[
  \begin{array}{c}
    x_{i1}(t) \\
    x_{i2}(t) \\
  \end{array}
\right] &
      =\left[
          \begin{array}{cc}
            0 & 1 \\
            -\csc \t & \cot \t \\
          \end{array}
        \right]\left[
  \begin{array}{c}
    x_{i1}(t-1) \\
    x_{i1}(t) \\
  \end{array}
\right]+\left[
          \begin{array}{c}
            0 \\
            u_i(t-1) \\
          \end{array}
        \right],
  \end{aligned}
\end{equation}
we propose the following encoder $\varphi_i$ for agent $i$:
\begin{equation}\label{encoder}
\left\{\begin{aligned}
          s_i(1)& =  q_t(\frac{y_i(1)}{p(0)}), ~
          \hat{x}_{i1}(1)=p(0)s_i(1); \\
          s_i(2)& = q_t(\frac{y_i(2)}{p(1)}), ~
          \hat{x}_{i1}(2)=p(1)s_i(2), \\
          \hat{x}_{i2}(2)& = \cot\t\hat{x}_{i1}(2)-\csc\t\hat{x}_{i1}(1); \\
          s_i(t)& = q_t(\frac{y_i(t)-[\cos\t\hat{x}_{i1}(t-1)
          +\sin\t\hat{x}_{i2}(t-1)]}{p(t-1)}),\\
          \hat{x}_{i1}(t)& = \cos\t\hat{x}_{i1}(t-1)+\sin\t\hat{x}_{i2}(t-1)+p(t-1)s_i(t), \\
          \hat{x}_{i2}(t)& = \cot\t\hat{x}_{i1}(t)-\csc\t\hat{x}_{i1}(t-1),\;t\geq 2,
\end{aligned}\right.\end{equation}
where $p(t)=p_0\g^t,0<\g<1$ is a decaying scaling function.

After $s_i(t)$ is received by one of the $i$-th agent's out-neighbors,
say $v\in \mathcal{N}_{i}^{-}$, a decoder $\varphi_{iv}$ will be activated:
\begin{equation}\label{decoder}
\left\{\begin{aligned}
          \hat{x}_{iv1}(1)& =p(0)s_i(1); \\
          \hat{x}_{iv1}(2)& =p(1)s_i(2), ~
          \hat{x}_{iv2}(2)=  \cot\t\hat{x}_{iv1}(2)-\csc\t\hat{x}_{iv1}(1); \\
          \hat{x}_{iv1}(t)& = \cos\t\hat{x}_{iv1}(t-1)+\sin\t\hat{x}_{iv2}(t-1)+p(t-1)s_i(t), \\
          \hat{x}_{iv2}(t)& = \cot\t\hat{x}_{iv1}(t)-\csc\t\hat{x}_{iv1}(t-1),\;t\geq 2.
\end{aligned}
\right.
\end{equation}

\begin{rem}
As in \cite{qiu_quantized_2016}, a scaled ``prediction error'' is quantized to generate the signal $s_i(t)$,
in an effort to reduce the number of quantization levels.
$s_i(t)$ is then used to construct the estimate $\hat{x}_{i1}(t)$ of the first component ${x}_{i1}(t)$,
which is combined with $\hat{x}_{i1}(t-1)$ to obtain the estimate $\hat{x}_{i2}(t)$ for ${x}_{i2}(t)$.
Denote $\Delta_i(t)=s_i(t)-d_i(t)$ as the quantization error, where
\begin{equation}\label{d(t)}
d_i(t)=\left\{\begin{array}{l}
 \frac{y_i(t)}{p(t-1)}, \; t=1,2; \\
\frac{y_i(t)-[\cos\t \hat{x}_{i1}(t-1)+\sin\t \hat{x}_{i2}(t-1)]}{p(t-1)}, ~
          t>2,
\end{array}\right.
\end{equation} and
$e_{ij}(t)=\hat{x}_{ij}(t)-{x}_{ij}(t)$
as the estimation for ${x}_{ij}(t),j=1,2$.
Then comparing (\ref{observer}), (\ref{encoder}) and (\ref{decoder}) we have
\begin{equation}\label{esti_err}
\left\{
\begin{array}{l}
  e_{i1}(t)= \hat{x}_{i1}(t)-{x}_{i1}(t)=p(t-1)\Delta_i(t),~t\geq 1; \\
  e_{i2}(t)= \hat{x}_{i2}(t)-{x}_{i2}(t)
  =\cot\t e_{i1}(t)-\csc\t e_{i1}(t-1)-u_{i}(t-1),~t\geq 2.
\end{array}
\right.
\end{equation}
Evidently the estimation error is related with control inputs
in addition to quantization errors, which may impair the consensus.
But as shown in the consensus analysis below, the influence of the control inputs can
be ignored by making the control gains arbitrarily small.
\end{rem}

Based on the outputs of the encoding-decoding scheme,
the distributed control law of agent $ i $ is given by
\begin{equation}\label{control1}
u_i(t)=\left\{
\begin{array}{l}
  0,\; t=0,1; \\
  \sum_{j=1}^{2}k_j\sum_{v\in
\mathcal{N}_{i}^{+}}g_{iv}[\hat{x}_{vij}(t)-\hat{x}_{ij}(t)], \;t\geq 2.
\end{array}
\right.
\end{equation}

\subsection{Consensus Analysis and Data Rate}
Some notations are defined as follows:
\begin{equation}\label{notation}
\begin{aligned}
u(t)&=[u_1(t),\dots,u_N(t)]^T,\\
\Delta(t)&=[\Delta_1(t),\dots,\Delta_N(t)]^T, \\
d(t)&=[d_1(t),\dots,d_N(t)]^T,\\
x_j(t)&=[x_{1j}(t),\dots,x_{Nj}(t)]^T,\\
\delta_j(t)&=(I_N-\phi_1\psi_{1}^T)x_{j}(t)
=[\delta_{1j}(t),\dots,\delta_{Nj}(t)],\\
e_j(t)&=[e_{1j}(t),\dots,e_{Nj}(t)]^T. \\
\end{aligned}
\end{equation}

We adopt the following two assumptions in the subsequent analysis.

\begin{assu}\label{ass1m=1}
The communication graph ${\mathcal{G}}$ contains a spanning tree.
\end{assu}

\begin{assu}\label{ass2m=1}
There exist known positive constants $C^\ast$ and $C_\delta^\ast$ such
that $\displaystyle \max_{j=1,2}||x_j(0)||\leq C^\ast$ and
$\displaystyle \max_{j=1,2}||\delta_j(0)||\leq
C_{\delta}^\ast$.
\end{assu}

\begin{rem}
Assumption \ref{ass1m=1} is a standard assumption, under which we have
$0<\text{Re}\l_2\le\dots\le\text{Re}\l_l$, with 0 as the simple eigenvalue.
Assumption \ref{ass2m=1} enables us to make the quantizer $q_t(\cdot)$ unsaturated at initial
steps.
\end{rem}

The following lemma is critical in the consensus analysis.

\begin{lem}\label{lemm=1}
Denote $K=\left[
            \begin{array}{cc}
              0 & 0 \\
              k_1 & k_2 \\
            \end{array}
          \right]
$ and $A_i=A-\l_iK$ with $\text{Re}\l_i>0$.
Let $k_j=c_j\v,j=1,2$ and $\v>0$. Then the following results hold with sufficiently small $\v$:

1). The spectral radius $\rho_i$ of $A_i$ is less than 1
  if $c_2\cos\t-c_1\sin\t>0$ and $c_1\cos\t+c_2\sin\t=0$.
  Moreover, $\rho_i=1-\frac{1}{2}(\text{Re}\l_i)(c_2\cos\t-c_1\sin\t)\v+o(\v)$.

2). Take $c_1,c_2$ as in 1). For any vector $\xi\in \R^2$, the entries of $A_{i}^s\xi$,
  which are denoted as $\xi_{s1}$ and $\xi_{s2}$, satisfy that $|\xi_{sj}|\le 5/2\rho_{i}^{s}$ for $j=1,2$.
\end{lem}

\begin{IEEEproof}
1). Noticing that $A=P\left[\begin{array}{cc}
                                 e^{\jmath\t} &  \\
                                  & e^{-\jmath\t}
                               \end{array}\right]P^{-1}$ with
                               $P=\left[
                                   \begin{array}{cc}
                                     1 & 1 \\
                                     \jmath & -\jmath \\
                                   \end{array}
                                 \right]$ and
                               $P^{-1}=\frac{1}{2}\left[
                                   \begin{array}{cc}
                                     1 & -\jmath \\
                                     1 & \jmath \\
                                   \end{array}
                                 \right]
$, we have
\begin{equation}\label{P}
\begin{aligned}
    \mu I-A& = P\left[\begin{array}{cc}
                                 \mu-e^{\jmath\t} &  \\
                                  & \mu-e^{-\jmath\t}
                               \end{array}\right]P^{-1} \\
     &= \frac{1}{2}\left[
     \begin{array}{cc}
     (\mu-e^{\jmath\t})+(\mu-e^{-\jmath\t}) &
     -\jmath(\mu-e^{\jmath\t})+\jmath(\mu-e^{-\jmath\t}) \\
     \jmath(\mu-e^{\jmath\t})-\jmath(\mu-e^{-\jmath\t}) &
     (\mu-e^{\jmath\t})+(\mu-e^{-\jmath\t}) \\
     \end{array}
     \right].
  \end{aligned}
\end{equation}
Consequently the characteristic polynomial of $A_i$ can be obtained as
\begin{equation}\label{chi}
\chi_i(\mu)=(\mu-e^{\jmath\t})(\mu-e^{-\jmath\t})
+\frac{\lambda_i}{2}\v[(c_2+c_1\jmath)(\mu-e^{\jmath\t})
+(c_2-c_1\jmath)(\mu-e^{-\jmath\t})].
\end{equation}
By perturbation theory \cite{seyranian_multiparameter_2003} it is readily seen that the two perturbed roots of \eqref{chi} are given by
\begin{equation}
\mu_{i1}=e^{\jmath\t}+\mu_{i11}\v+o(\v),~
\mu_{i2}=e^{-\jmath\t}+\mu_{i21}\v+o(\v).
\end{equation}
Substituting $\mu=\mu_{i1}$ into $\chi_i(\mu)=0$ and comparing the coefficient of $\v$ yield
$$
\mu_{i11}(2\jmath\sin\t)+\frac{1}{2}\lambda_i(c_2-c_1\jmath)(2\jmath\sin\t)=0
$$
and $\mu_{i11}=-\frac{1}{2}\lambda_i(c_2-c_1\jmath)$ follows immediately.
Direct computation shows that $|\mu_{i1}|^2=1+2\text{Re}(\mu_{i11}e^{-\jmath\t})\v+o(\v)$,
where
$$\text{Re}(\mu_{i11}e^{-\jmath\t})=-\frac{1}{2}[a_i(c_2\cos\t-c_1\sin\t)+b_i(c_1\cos\t+c_2\sin\t)]$$
if we let $\l_i=a_i+b_i\jmath$.
Clearly $|\mu_{i1}|=1-\frac{1}{2}(\text{Re}\l_i)(c_2\cos\t-c_1\sin\t)\v+o(\v)$
when $c_1\cos\t+c_2\sin\t=0$.
Similarly we can show $\mu_{i21}=-\frac{1}{2}\lambda_i(c_2+c_1\jmath)$
and $|\mu_{i2}|=1-\frac{1}{2}\text{Re}\l_i(c_2\cos\t-c_1\sin\t)\v+o(\v)$,
which implies the conclusion.

2). Here we need to compute the Jordan decomposition of $A_i$.
The eigenvector corresponding to the eigenvalue $\mu_{i1}$ is given by
$w_{i1}=w_{i10}+w_{i11}\v+o(\v)$.
Substituting it into the equation $A_iw_{i1}=\mu_{i1}w_{i1}$ and comparing the coefficients of constant term,
we have $Aw_{i10}=e^{\jmath\t}w_{i10}$. With the normalization condition $v^Tw_{i1}=1$
where $v^T=\frac{1}{2}(1~-\jmath)$, $w_{i10}=(1~\jmath)^T$.
Similarly, the eigenvector corresponding to the eigenvalue $\mu_{i2}$ is given by
$w_{i2}=w_{i20}+w_{i21}\v+o(\v)$ with $w_{i20}=(1~-\jmath)^T$.
Letting $R_i=(w_{i1}~w_{i2})=\left[
                               \begin{array}{cc}
                                 1+O(\v) & 1+O(\v) \\
                                 \jmath+O(\v) & -\jmath+O(\v) \\
                               \end{array}
                             \right]
$, it is clear that $R_{i}^{-1}=\frac{1}{\det R_i}\left[
                               \begin{array}{cc}
                                 -\jmath+O(\v) & -1+O(\v) \\
                                 -\jmath+O(\v) & 1+O(\v) \\
                               \end{array}
                             \right]=\frac{1}{2}\left[
                                   \begin{array}{cc}
                                     1+O(\v) & -\jmath+O(\v) \\
                                     1+O(\v) & \jmath+O(\v) \\
                                   \end{array}
                                 \right]$.
The result follows directly by noticing that
$A_i=R_i\left[
          \begin{array}{cc}
            \mu_{i1} &  \\
             & \mu_{i2} \\
          \end{array}
        \right]R_{i}^{-1}.
$ \end{IEEEproof}

\begin{rem}\label{cj1}
Denote $\rho=\max\limits_{i=2,\dots,l}\rho_i$
and let $h$ be a constant in $(0,\text{Re}\lambda_2]$.
Taking $c_1=-\sin\t/h$ and $c_2=\cos\t/h$,
we have $\rho\le 1-\v+o(\v)<1-\v/2$ with sufficiently small $\v$.
\end{rem}

We also need to define some constants as follows:
\begin{equation}\label{constant}
\begin{aligned}
  C_0 &= \frac{1}{2}|c_1|+\frac{3}{2}|c_2\csc\t|=\frac{1}{2h}(|\sin\t|+3|\cot\t|), \\
  \Lambda& = \max_{i=2,\dots,l}|\l_i|,\\
  C(1)&= ||U_{L}^{-1}||+2C_0\Lambda||U_{L}||,\\
  C(k)&= ||U_{L}^{-1}||+2C_0(\Lambda+1)||U_{L}||+10(|c_1|+|c_2|)C(k-1),~k\ge 2,\\
  \bar C&= 5(|c_1|+|c_2|)C(N_{\max})+C_0||U_{L}||,
\end{aligned}
\end{equation}
where $N_{\max}=\max_{i=2,\dots,l}N_l$.

\begin{lem}\label{epsm=1}
Let $\g=1-\v/4$. Then we can choose sufficiently small
$\v$ to satisfy the following inequalities:
\begin{subequations}
\begin{align}
(\Lambda+1)\bar C\v\le \frac{1}{2}\g|\csc\t|||U_{L}||;\label{eps11}\\
\frac{1}{\g}(2|\cos\t|+\frac{1}{\g})\le 2|\cos\t|+1+\frac{1}{2};\label{eps12}\\
(N-1)\bar C(\Lambda+1)\v\le\frac{1}{4}|\csc\t|\g^3.\label{eps13}
\end{align}
\end{subequations}
\end{lem}

\begin{thm}\label{consensus&rate}
Take $c_j$'s as in Remark \ref{cj1} and let $\g=1-\v/4$.
Select sufficiently small $\v$ to satisfy Lemmas \ref{lemm=1} and \ref{epsm=1}
with $\rho<1-\v/2$. Then under Assumptions \ref{ass1m=1}
and \ref{ass2m=1}, consensus can be achieved at a convergence rate of $O(\g^t)$
provided that $g_0\geq \max\{\frac{4}{3\g}C^*,C_{\delta}^*\}$ and $M(t)$ satisfies
\begin{equation}\label{M(t)m=1}
  \left\{\begin{array}{l}
          M(t)\geq 1, \; t=1,2; \\
          M(t)\geq |\cos\t|+1/2, ~ t=2m+1,\dots.
        \end{array}
  \right.
\end{equation}
Therefore, the number of bits used to achieve
the consensus is $\lceil \log_2 2\lceil
|\cos\t|+1/2\rceil \rceil$.
\end{thm}

\begin{IEEEproof}
1) Preparation. The closed-loop system of disagreement vectors can be established as
$$
\left[
  \begin{array}{c}
    \delta_{1}(t+1) \\
    \delta_{2}(t+1) \\
  \end{array}
\right]=\left[
          \begin{array}{cc}
            \cos \t I_N & \sin \t I_N \\
            -\sin \t I_N & \cos \t I_N \\
          \end{array}
        \right]\left[
  \begin{array}{c}
    \delta_{1}(t) \\
    \delta_{2}(t) \\
  \end{array}
\right]+\left[
  \begin{array}{c}
    0 \\
    u(t)
    \end{array}\right]
$$
with
\begin{equation}\label{controllaw1}
  u(t)=\left\{\begin{array}{l}
                0, ~t=0,1;  \\
                -\sum_{j=1}^{2}k_jL(\delta_j(t)+e_j(t)),~ t\geq 2,
              \end{array}
  \right.
\end{equation}
by noticing (\ref{control1}) and $L=L(I_N-\phi_1\psi_{1}^T)$.
Letting $\tilde{\delta}_j(t)
=U_{L}^{-1}\delta_j(t)=
[\tilde{\delta}_{1j}(t),\dots,\tilde{\delta}_{N,j}(t)]^T$,
we obtain
$$
\left[
  \begin{array}{c}
    \tilde{\delta}_{1}(t+1) \\
    \tilde{\delta}_{2}(t+1) \\
  \end{array}
\right]=\left[
          \begin{array}{cc}
            \cos \t I_N & \sin \t I_N \\
            -\sin \t I_N-k_1L_J & \cos \t I_N-k_2L_J  \\
          \end{array}
        \right]\left[
  \begin{array}{c}
    \tilde{\delta}_{1}(t) \\
    \tilde{\delta}_{2}(t) \\
  \end{array}
\right]+\left[
  \begin{array}{c}
    0 \\
    \eta(t)
    \end{array}\right],
$$
where $\eta(t)=-L_JU_{L}^{-1}(k_1e_1(t)+k_2e_2(t))$.
Denote
$\tilde{\delta}^{i}(t)=[\tilde{\delta}_{i1}(t),\tilde{\delta}_{i2}(t)]^T$ for $i=1,\dots,N$.
Clearly $\tilde{\delta}^{1}(t)\equiv 0$ due to that
$\tilde{\delta}_{1j}(t)=\psi_{1}^T(I_N-\phi_1\psi_{1}^T)x_j(t)=0$ for $j=1,2$.
Without loss of generality we assume that $N_2=2$ (the Jordan block with respect to $ \l_2 $ is two-dimensional)
and consequently $\tilde{\delta}^{2}(t)$ and $\tilde{\delta}^{3}(t)$
are coupled in the following way:
\begin{equation}\label{disagreement}
\begin{aligned}
  \tilde{\delta}^{i}(t+1)& = A\tilde{\delta}^{i}(t),~t=0,1,~i=2,3; \\
  \tilde{\delta}^{2}(t+1)& = A_2\tilde{\delta}^{2}(t)-K\tilde{\delta}^{3}(t)-\eta_2(t), \\
  \tilde{\delta}^{3}(t+1)& = A_2\tilde{\delta}^{3}(t)-\eta_3(t),~t\ge 2,
\end{aligned}
\end{equation}
where $A_2$ and $K$ have been defined in Lemma \ref{lemm=1}
and $\eta_2(t)=[0,(\l_2\psi_{2}^T+\psi_{3}^T)(k_1e_1(t)+k_2e_2(t))]^T$,
$\eta_3(t)=[0,\l_2\psi_{3}^T(k_1e_1(t)+k_2e_2(t))]^T$.

2) Estimation error and exponential convergence.
Remember that $e_2(t)=\cot\t e_1(t)-\csc\t e_1(t-1)-u(t-1)$ is dependent on the control input
by \eqref{esti_err},
we have to first make an estimate for $u(t)$ before establishing the consensus result.
Below we shall show $|\psi_{i}^{T}u(t)|\leq\v(\Lambda+1)\bar Cp_0\g^{t-2},t\geq 2$
for $i\ge 2$ by induction.

With the choice of $p_0$ and $\g$ it is easy to see $|s_i(t)|\leq 3/2$
when $t\leq 2$ by noticing $|y_i(t)|\leq 2C^\ast$, hence we obtain
$\max\limits_{t=1,2}||\Delta(t)||\leq 1/2$ if $M(1),M(2)\ge 1$.
For $i=2$, we have $\psi_{2}^{T}L=\l_2\psi_{2}^{T}+\psi_{3}^{T}$
and as a result
\begin{equation*}
\begin{aligned}
|\psi_{2}^{T}u(2)|& = |\psi_{2}^TL\sum_{j=1}^2 k_j(\delta_j(2)+e_j(2))| \\
& \le \v(\Lambda+1)||U_{L}||\big[|c_1|(2||\delta(0)||+p(1)||\Delta(2)||)\\
&+|c_2|(2||\delta(0)||+|\cot\t|p(1)||\Delta(2)||+|\csc\t|p(0)||\Delta(1)||)\big]\\
& \le \v (\Lambda+1)||U_{L}||(|c_1|+|c_2|)(2||\delta(0)||+|\csc\t|p_0) \\
& \le \v (\Lambda+1)||U_{L}||(|c_1|+|c_2|)(2+|\csc\t|)p_0\\
& \le \v (\Lambda+1)\bar Cp_0,
\end{aligned}
\end{equation*}
which also holds for $|\psi_{i}^{T}u(2)|$ for $i> 2$.

Now assume that
\begin{align}\label{induction}
\begin{aligned}
|\psi_{i}^{T}u(t)| \leq& \v(\Lambda+1)\bar Cp_0\g^{t-2},~t\ge 2;\\
  ||\Delta(\tau)|| &\le1/2 (\Rightarrow |e_{i1}(\tau)|\le \frac{1}{2}p_0\g^{\tau-1}),\; 1\leq \tau\leq t.
\end{aligned}
\end{align}
Then by combining \eqref{induction} and \eqref{eps11} it follows that
\begin{equation}\label{eta3}
\begin{aligned}
  ||\eta_3(\tau)|| &= |\l_2||\psi_{3}^T(k_1e_1(\tau)+k_2e_2(\tau))| \\
  &\le  \v|\l_2|p_0\g^{\tau-2}||U_{L}||(\frac{1}{2}|c_1|+|c_2||\csc\t|)
  +\v|\l_2|c_2||\psi_{3}^Tu(\tau-1)|\\
  &\le  \v|\l_2|||U_{L}||C_0p_0\g^{\tau-2}.
\end{aligned}
\end{equation}
Recalling \eqref{disagreement} we get that for $t\ge2$
\begin{align*}
\tilde{\delta}^3(t+1)=A_{2}^{t-1}\tilde{\delta}^3(2)-\sum_{\tau=1}^{t-1}A_{2}^{t-1-\tau}\eta_3(\tau+1),
\end{align*}
which produces the following estimate by Lemma \ref{lemm=1} and \eqref{eta3}
\begin{equation}
\begin{aligned}
  ||\tilde{\delta}^3(t+1)||& \le \frac{5}{2}(\rho_{2}^{t-1}||\tilde{\delta}^3(2)||
  +4|\l_2|C_0||U_{L}||p_0\g^{t-1}) \\
  & \le 5\g^{t-1}(||U_{L}^{-1}||||\delta(0)||+2|\l_2|C_0||U_{L}||p_0) \\
  & \le 5C(1)p_0\g^{t-1}.
\end{aligned}
\end{equation}
Similarly, an estimate for $||\tilde{\delta}_2(t+1)||$ can be found as
$
||\tilde{\delta}^2(t+1)||\le 5C(2)p_0\g^{t-1}
$,
if we notice that $||K\tilde{\delta}^3(\tau)||\le 5\v(|c_1|+|c_2|) C(1)p_0\g^{\tau-2}$
and $||\eta_2(\tau)||\le \v (|\l_2|+1)||U_{L}||C_0p_0\g^{\tau-2}$ for $2\le \tau\le t$.
For any $i\ge 2$, by proceeding along the same line as in the above it is concluded that
\begin{equation}\label{esti_disagree}
||\tilde{\delta}^i(t+1)||\le 5C(N_{\max})p_0\g^{t-1}.
\end{equation}

3) Data rate. Now we are able to discuss the estimation for
$|\psi_{i}^{T}u(t+1)|$, which is bounded by the sum of
$|\sum_{j=1}^2k_j\psi_{i}^{T}L\delta_j(t+1)|$ and
$|\sum_{j=1}^2k_j\psi_{i}^{T}Le_j(t+1)|$. For the first term, by
\eqref{esti_disagree} it is readily seen that
\begin{equation}\label{tem1}
\begin{aligned}
   |\sum_{j=1}^2k_j\psi_{i}^{T}L\delta_j(t+1)|
  &\le |\sum_{j=1}^2k_j\psi_{i}^{T}U_{L}L_J\tilde\delta_j(t+1)|\\
  &\le  5\v(|c_1|+|c_2|)C(N_{\max})p_0\g^{t-1}; \\
\end{aligned}
\end{equation}
while the second term is essentially related with $e_j(t+1)$, or more exactly $\Delta(t+1)$.
By \eqref{d(t)} and \eqref{esti_err} we have
\begin{equation}\label{d(t+1)}
\begin{aligned}
  d(t+1)& = \frac{1}{p(t)}(-\cos\t e_1(t)-\sin\t e_2(t)) \\
  & = \frac{-2\cos\t}{\g}\Delta(t)+\frac{1}{\g^2}\Delta(t-1)+\frac{1}{p(t)}\sin\t u(t-1),
\end{aligned}
\end{equation}
which is obviously dependent on the previous quantization errors $\Delta(t)$ and $\Delta(t-1)$,
as well as the previous control input $u(t-1)$. Hence with the induction assumption \eqref{induction}
the quantizer can be made unsaturated with sufficiently many bits at time $t+1$,
and $||\Delta(t+1)||\le 1/2$ follows directly.
Consequently
\begin{align}\label{tem2}
|\sum_{j=1}^2k_j\psi_{i}^{T}Le_j(t+1)|\le \v|\l_2|||U_{L}||C_0p_0\g^{t-1}
\end{align}
as in \eqref{eta3}.
The induction is then established by combining \eqref{tem1}. Moreover,
by \eqref{esti_disagree} the consensus can be achieved at a
convergence rate of $O(\g^t)$.

Below we are to calculate the number of required quantization levels at each time step.
The situation when $t\leq 2$ has been discussed.
When $t>2$, from \eqref{d(t+1)} we can see that
\begin{align*}
  ||d(t)||& \le  \frac{1}{2\g}(2|\cos\t|+\frac{1}{\g})+\frac{|\sin\t|}{p(t-1)} ||u(t-2)||\\
  & \le  \frac{1}{2}(2|\cos\t|+1)+\frac{1}{4}+\frac{|\sin\t|}{\g^3}(N-1)(\Lambda+1)\bar C\\
  & \le  \frac{1}{2}(2|\cos\t|+1)+\frac{1}{2}
\end{align*}
by noticing \eqref{eps12}, \eqref{eps13} and $u(t)=\sum_{i=2}^{N}\phi_{i}\psi_{i}^{T}u(t)$ ($\psi_{1}^{T}u(t)=0$).
In summary, the proof is completed. \qquad\end{IEEEproof}

\begin{rem}
For the coupling system shown in \eqref{disagreement},
we divide it into two subsystems with disturbance.
Each subsystem can be stabilized as long as the disturbance decays exponentially
at a speed slower than $\rho_2$, i.e. $||\eta_3(t)||\sim O(\g^t)$ and
$||K\tilde{\delta}^{3}(t)+\eta_2(t)||\sim O(\g^t)$, with $\rho_2<\g<1$.
The interference of $u(t)$ in the estimation error $e(t)$ can be ignored,
as long as $||u(t)||\sim O(\v^\alpha)p_0\g^t$ with $\alpha>0$,
yielding that $||\eta_3(t)||\sim O(\v)p_0\g^t$, and then $||\tilde{\delta}^{3}(t)||\sim O(\g^t)$ follows.
As a result, $||K\tilde{\delta}^{3}(t)||\sim O(\v)\g^t$ and $||\tilde{\delta}^{3}(t)||\sim O(\g^t)$
follows by combining $||\eta_3(t)||\sim O(\v)p_0\g^t$.
Such a reasoning still applies when \eqref{disagreement} involves more than two subsystems.
Finally we show that $||u(t)||\sim O(\v)p_0\g^t$,
and by \eqref{d(t+1)} we conclude that the control input does not consume extra bits
in exchanging the information when the control gains are sufficiently small.
\end{rem}

\section{Higher-order cases}\label{sec:high order}

In this section, we will conduct the same task as in the last section
for general higher-order cases.
The analysis actually proceeds along a similar line,
but the assignment of control gains to achieve consensus is much more challenging,
and we have to resort to combinatorial identities for an explicit data rate.
As before, we first provide an encoding-decoding scheme for all the
agents and devise a control protocol in terms of the outputs of the scheme.
Then we present some lemmas, which will play a crucial role
in the convergence analysis and the derivation of the data rate in the final part.

\subsection{Encoding-decoding scheme and distributed control law}

As pointed out in the last section, the construction of
the encoding scheme should follow two principles: firstly,
the encoder is able to estimate other state variables
given that only the first component is measurable;
secondly, the estimation should be based on iterations in an effort to reduce quantization levels.
Such an idea can be stated more clearly as follows.
At each time step, the scaled difference between the
output $y_i(t)$ and its estimate is quantized to obtain a
signal $s_i(t)$. Based on $s_i(t)$ we construct an estimate
$\hat{x}_{i1}(t)$ of the first component ${x}_{i1}(t)$, and
combine previous estimates $\hat{x}_{i1}(t-1)$ through
$\hat{x}_{i1}(t-2m+1)$ to obtain estimates of the other components
${x}_{i2}(t)$ through ${x}_{i,2m}(t)$.

To be detailed, denote the observability
matrix $\mathcal {O}=\left[
         \begin{array}{c}
         I_{2m}(1,\cdot) \\
         A(1,\cdot) \\
         \vdots \\
         A^{2m-1}(1,\cdot) \\
         \end{array}
         \right]$,
\begin{align*}
 \bar x_{i}(t) &= [x_{i1}(t-2m+1), x_{i1}(t-2m+2), \dots,
x_{i1}(t)]^T, \\
b_n(\t) &= [0, \dots, 0, A(1,2m), \dots, A^{n-1}(1,2m)]^T\in
\R^{2m}.
\end{align*}
 We have
\begin{equation}
\bar x_{i}(t)=\mathcal {O}x^i(t-2m+1)+\sum_{n=1}^{2m-1}b_n(\t)
u_{i}(t-n)
\end{equation}
if we notice by (\ref{conjudynamics}) that
\begin{equation}
\begin{aligned}
  x^{i}(t-2m+k)&= A^{k-1}x^{i}(t-2m+1)+\sum\limits_{n=0}^{k-2}A^{k-2-n}bu_{i}(t-2m+1+n),\\
  & k=1,\dots,2m.
\end{aligned}
\end{equation}
As a result,
\begin{equation}x^i(t-2m+1)=\mathcal
{O}^{-1}[\bar x_{i}(t)-\sum_{n=1}^{2m-1}b_n(\t) u_{i}(t-n)]
\end{equation} and
\begin{equation}\label{realstates}
\begin{aligned}
  x^i(t)& = A^{2m-1}x^i(t-2m+1)
  +\sum_{j=0}^{2m-2}A^{2m-2-j}bu_{i}(t-2m+1+j) \\
  & = S\bar x_{i}(t)+\sum_{j=1}^{2m-1}\tilde{b}_j(\t)u_{i}(t-j),
\end{aligned}
\end{equation}
where $S=A^{2m-1}\mathcal {O}^{-1}$
(the existence of $\mathcal {O}^{-1}$ can be easily verified by PBH test \cite{chen_linear_1995} if
$\sin\t\neq 0$) and $\tilde{b}_n(\t)=-S{b}_n(\t)+A^{n-1}(\cdot,2m)$.
Inspired by \eqref{realstates}, the encoding scheme for agent $i$ is implemented below:

\noindent for $t\le 2m$,
\begin{equation}\label{encoder1}
\left\{
\begin{array}{l}
  s_i(t)=  q_t(\frac{y_i(t)}{p(t-1)}), \;
  \hat{x}_{i1}(t)=  p(t-1)s_i(t);\\
  \left[
    \begin{array}{c}
      \hat{x}_{i2}(2m) \\
      \vdots \\
      \hat{x}_{i,2m}(2m) \\
    \end{array}
  \right]= S_m\left[
    \begin{array}{c}
      \hat{x}_{i1}(1) \\
      \vdots \\
      \hat{x}_{i1}(2m) \\
    \end{array}
  \right];
\end{array}
\right.\\
\end{equation}
for $t> 2m$,
\begin{equation}\label{encoder2}
\left\{
\begin{array}{l}
  s_i(t)= q_t(\frac{y_i(t)-[\cos\t \hat{x}_{i1}(t-1)+\sin\t \hat{x}_{i2}(t-1)
          +\hat{x}_{i3}(t-1)]}{p(t-1)}), \\
  \hat{x}_{i1}(t)= \cos\t \hat{x}_{i1}(t-1)+\sin\t \hat{x}_{i2}(t-1)+\hat{x}_{i3}(t-1)+p(t-1)s_i(t),\\
  \left[
    \begin{array}{c}
      \hat{x}_{i2}(t) \\
      \vdots \\
      \hat{x}_{i,2m}(t) \\
    \end{array}
  \right]= S_m\left[
    \begin{array}{c}
      \hat{x}_{i1}(t-2m+1) \\
      \vdots \\
      \hat{x}_{i1}(t) \\
    \end{array}
  \right],\\
\end{array}
\right.
\end{equation}
where $S_m=S(2:2m,\cdot)$ is a submatrix of $S$ obtained by deleting the
first row, and $p(t)=p_0\g^t,0<\g<1$ is a decaying scaling function.

After $s_i(t)$ is generated, transmitted and received by one
of agent $i$'s out-neighbors, say $v\in \mathcal{N}_{i}^{-}$, a decoder will be activated:

\noindent for $t\le 2m$,
\begin{equation}\label{decoder1}
\left\{\begin{array}{l}
         \hat{x}_{iv1}(t)= p(t-1)s_i(t); \\
         \left[
    \begin{array}{c}
      \hat{x}_{iv2}(2m) \\
      \vdots \\
      \hat{x}_{iv,2m}(2m) \\
    \end{array}
  \right]= S_m\left[
    \begin{array}{c}
      \hat{x}_{iv1}(1) \\
      \vdots \\
      \hat{x}_{iv1}(2m) \\
    \end{array}
  \right];
\end{array}
\right.
\end{equation}
for $t> 2m$,
\begin{equation}\label{decoder2}
\left\{\begin{array}{l}
\hat{x}_{iv1}(t)=  \cos\t
\hat{x}_{iv1}(t-1)+\sin\t\hat{x}_{iv2}(t-1)+\hat{x}_{iv3}(t-1)+p(t-1)s_i(t);\\
\left[
    \begin{array}{c}
      \hat{x}_{iv2}(t) \\
      \vdots \\
      \hat{x}_{iv,2m}(t) \\
    \end{array}
  \right]=  S_m\left[
    \begin{array}{c}
      \hat{x}_{iv1}(t-2m+1) \\
      \vdots \\
      \hat{x}_{iv1}(t) \\
    \end{array}
  \right].\\
       \end{array}
\right.
\end{equation}

\begin{rem}
Comparing (\ref{encoder1}) with (\ref{decoder1}), (\ref{encoder2}) with (\ref{decoder2}),
it is clear that
$
\hat{x}_{ivj}(t)\equiv \hat{x}_{ij}(t),~j=1,\dots,2m,
$
for $v\in \mathcal{N}_{i}^{-},i=1,\dots,N$.
Denote $e_{ij}(t)=\hat{x}_{ij}(t)-{x}_{ij}(t)$ as the estimation error,
$\Delta_i(t)=s_i(t)-d_i(t)$ as the quantization error, where
\begin{equation}\label{di(t)m2}
d_i(t)=\left\{\begin{array}{l}
\frac{y_i(t)}{p(t-1)}, \; t=1,2,\dots,2m; \\
\frac{y_i(t)-[\cos\t \hat{x}_{i1}(t-1)+\sin\t \hat{x}_{i2}(t-1)
          +\hat{x}_{i3}(t-1)]}{p(t-1)},\;t>2m. \\
\end{array}\right.
\end{equation}
Comparing (\ref{realstates}) with
(\ref{encoder2}), the estimation errors are given by the following:
\begin{equation}\label{esti_errh}
\left\{
\begin{aligned}
  e_{i1}(t)& = p(t-1)\Delta_i(t),~t\ge 1; \\
  e_{ij}(t)& = \sum_{n=1}^{2m}S(j,n)e_{i1}(t-2m+n)
  -\sum_{n=1}^{2m-1}\tilde{b}_{nj}(\t)u_{i}(t-n),\\
  &t\geq 2m,j=2,\dots,2m,
\end{aligned}
\right.
\end{equation}
where $\tilde{b}_{nj}$ is the $j$-th entry of $\tilde{b}_{n}$.
\end{rem}

\begin{rem}\label{rem:coder comp}
	The encoding schemes \eqref{encoder} and \eqref{encoder2} proposed in our work is different 
	from those in \cite{meng_coordination_2015} or \cite{qiu_quantized_2016}.
	Actually, to address the general dynamics with unmeasurable states,
	\cite{meng_coordination_2015} designed the encoding scheme respectively for the output and control input, and used Luenberger observer to estimate the unmeasurable states. 
	If we compare with \cite{qiu_quantized_2016},
	we can also see a big difference: the special structure of $ n $-th order integrator dynamics enables it to easily ``recover'' the control input at $ n $ steps earlier, 
	based on which an estimate of the unmeasurable components can be made with time delay,
	and the encoding scheme can be designed accordingly.
	However, in our case it is unlikely to
	achieve the same task and we resort to the constructability of the system,
	namely we estimate the unmeasurable states directly from $\hat x_{i1}(t)$ through $\hat x_{i1}(t-2m+1)$.
	Although such a method introduces the control input into the estimation errors,
	it is able to make an estimation without time delay, 
	and hence avoids the stabilization of a time-delayed
	closed-loop system in the consensus analysis.
\end{rem}

For agent $i$, the outputs of encoder are $\hat{x}_{i1}(t),\dots,\hat{x}_{i,2m}(t)$,
while the outputs of decoders are $\hat{x}_{vi1}(t),\dots,\hat{x}_{vi,2m}(t)$ for $v\in \mathcal{N}_{i}^{+}$.
Based on these outputs, the distributed control law of agent $ i $ is proposed as
\begin{equation}\label{controlh}
u_i(t)=\left\{
\begin{array}{l}
  0,\; t=0,1,\dots,2m-1; \\
  \sum_{j=1}^{2m}k_j\sum_{v\in
\mathcal{N}_{i}^{+}}g_{iv}[\hat{x}_{vij}(t)-\hat{x}_{ij}(t)], \;t\geq 2m.
\end{array}
\right.
\end{equation}

\subsection{Lemmas}

The following two lemmas are respectively needed in analyzing consensus and data rate.
The first one is to stabilize the closed-loop system of disagreements,
and the second one is used for estimating the magnitude of $u_i(t)$ and $d_i(t)$.

\begin{lem}\label{lem1}
Denote $A_i=A-\lambda_iK$ with $\lambda_i>0$,
where $K\in \R^{2m\times 2m}$ and its
nonzero entries are only at the last row $[k_1, k_2, \dots, k_{2m-1}, k_{2m}]$.
Take
\begin{align}\label{kj}
\begin{aligned}
 k_{2j-1}&= \left\{\begin{array}{l}
  c_{2j-1}\v^{m-j}, ~ j=1,\dots,m-1; \\
  c_{2j-1}\v, ~j=m,
  \end{array}\right.\\
  k_{2j}&= \left\{\begin{array}{l}
  c_{2j}\v^{m-j}, ~j=1,\dots,m-1; \\
  c_{2j}\v, ~j=m.
  \end{array}
\right.
\end{aligned}
\end{align}
Then we can find constants $c_{2j-1}$ and $c_{2j}(j=1,\dots,m)$ such
that, when $\v$ is sufficiently small, the
spectral radius $\rho_i$ of $A_i$ is less than 1 with distinct eigenvalues.
Moreover, denote
\begin{equation}\label{H}
R_m=\frac{1}{2}+\frac{1}{2}(c_{2m-1}\sin\t-c_{2m}\cos\t),~
H=Re[\frac{c_{2m-5}\jmath-c_{2m-4}}{c_{2m-3}\jmath-c_{2m-2}}e^{-\jmath\t}].
\end{equation}
The requirements about
$c_{2j-1}$'s and $c_{2j}$'s corresponding to different $m$'s
are listed below.

1). $m=2$:
  let $c_{1}=-\sin2\t$ and $c_{2}=\cos2\t$. If
$R_2<0$, then
$\rho_i=1+\frac{1}{2}\lambda_iR_2\v+o(\v)$;

2). $m\geq 3$:
  let $c_{2m-3}=-\sin2\t$ and
  $c_{2m-2}=\cos2\t$.
 If $\lambda_iR_m+H<0$ and
$\text{Re}(\vartheta_{n1}e^{-\jmath\t})<0$ with $\vartheta_{n1},n=3,\dots,m$
denoting the $m-2$ distinct roots of the equation
\begin{equation}
\begin{array}{l}\label{eqn}
  \quad\vartheta_{1}^{m-2}(c_{2m-2}-c_{2m-3}\jmath)
  +\cdots+\vartheta_{1}(c_{4}-c_{3}\jmath)+(c_{2}-c_{1}\jmath)=0,
\end{array}
\end{equation}
then
$\rho_i=1+\frac{1}{2}\max_{n=3,\dots,n}
\{\lambda_iR_m+H,2\text{Re}(\vartheta_{n1}e^{-\jmath\t})\}\v+o(\v)$.
\end{lem}

\begin{lem}\label{lem2}
Assume that Lemma \ref{lem1} holds. When $\v$ is sufficiently small,
for any vector $\xi\in \R^{2m}$, the
entries of $A_{i}^{s}\xi$, which are denoted as $\xi_{s,2j-1}$ and
$\xi_{s,2j},j=1,\dots,m$, satisfy that
  \begin{equation}\label{entry_Ai}
  |\xi_{s,2j-1}|,|\xi_{s,2j}|\leq\left\{
  \begin{array}{l}
    ||\xi||M_{ij}\rho_{i}^s\v^{j-(m-1)},
    ~ j=1,2,\dots,m-2; \\
    ||\xi||M_{ij}\rho_{i}^s\v^{(j-m)/2},
    ~ j=m-1,m,
  \end{array}\right.
  \end{equation}
where
$$
M_{ij}=\left\{
\begin{array}{ll}
  \frac{5}{2\lambda_i}(\sum_{n=3}^{m}
  \frac{|\vartheta_{n1}|^{j-1}}{\prod\limits_{3\leq k\leq m,k\neq n}|\vartheta_{k1}-\vartheta_{n1}|}), & j=1,\dots,m-3; \\
  \frac{5}{2\lambda_i}(\sum_{n=3}^{m}
  \frac{|\vartheta_{n1}|^{m-3}}{\prod\limits_{3\leq k\leq m,k\neq n}|\vartheta_{k1}-\vartheta_{n1}|}+1), & j=m-2; \\
  \frac{3}{\sqrt{2\lambda_i}},\;j=m-1; \\
  5/2,\;j=m.
\end{array}
\right.
$$
\end{lem}

\begin{rem}
The proofs of the above lemmas can be found in the Appendix. As in \cite{qiu_quantized_2016}, the
basic idea is to combine the bifurcation analysis of the roots of
characteristic polynomials and the Jordan basis of a perturbed
matrix \cite{baumgartel_analytic_1985}. However, the situation here is much different.
On one hand, the complex conjugate eigenvalues of the original matrix $A$
complicates the analysis of the perturbed eigenvalues, as seen from the proof of Lemma \ref{lem1}.
On the other hand, unlike \cite{qiu_quantized_2016} where the unperturbed
matrix admits multiple eigenvalues of 0 and 1,
the unperturbed matrix here admits eigenvalues of $m$ identical pairs of complex conjugate numbers,
which allows a less cumbersome calculation of the perturbed Jordan basis,
as in the proof of Lemma \ref{lem2}.
\end{rem}

\begin{rem}\label{cj}
Assume $0< \lambda_2\leq\dots\leq \lambda_N$
and let $\rho=\max\limits_{i=2,\dots,N}\rho_i$, $h\in(0,\lambda_2]$.
Given $c_{2m-3}=-\sin2\t$ and $c_{2m-2}=\cos2\t$,
the other constants $c_{2j-1}$ and $c_{2j}$ can
be selected as follows such that $\rho\le 1-\v+o(\v)<1-\v/2$
holds with sufficiently small $\v$:

 1). $m=2$: select $c_3=-(4/h+1)\sin\t$,
  $c_4=(4/h+1)\cos\t$ such that $R_2=-2/h<0$;

 2). $m\geq 3$: first select
$c_{2m-4},c_{2m-5},\cdots,c_{1}$ such that the solutions of \eqref{eqn} are given by
$\vartheta_{n1}=-(n-2)e^{\jmath\t},n=3,\dots,m$ and $H$ is determined by \eqref{H}.
In fact, direct computation shows that
$c_{2m-4}-c_{2m-5}\jmath=\frac{1}{2}(m-1)(m-2)e^{3\t\jmath}$
and consequently $H=(m-1)(m-2)/2> 0$. Now
let $c_{2m-1}=-[(2H+4)/h+1]\sin\t$,
$c_{2m}=[(2H+4)/h+1]\cos\t$ such that
$$\lambda_iR_m+H=H(1-\lambda_i/h)-2\lambda_i/h<-2\lambda_i/h\le-2.$$
With such a selection, $M_{ij}=\frac{5}{2\lambda_i}(\sum_{n=1}^{m-2}
  \frac{n^{j-1}}{\prod\limits_{1\leq k\leq m-2,k\neq n}|k-n|}),\, j=1,\dots,m-3$
and $M_{i,m-2}=\frac{5}{2\lambda_i}(\sum_{n=1}^{m-2}
  \frac{n^{m-3}}{\prod\limits_{1\leq k\leq m-2,k\neq n}|k-n|}+1)$.
\end{rem}

To explicitly express the data rate, another lemma is required.

\begin{lem}\label{lem3}
Denote
$$
    l(\t) = \left[
              l_0(\t), l_1(\t), \dots, l_{2m-2}(\t), l_{2m-1}(\t)
          \right]  =
       \cos\t S(1,\cdot)+\sin\t S(2,\cdot)+S(3,\cdot).
$$ Then
$$
\begin{array}{l}
  l_k(\t)= (-1)^{k-1}\sum\limits_{h=0}^{\lfloor k/2\rfloor}
  C(m,k-2h)C(m-(k-2h),h)(2\cos\t)^{k-2h}, \;k=0,1,\dots,2m-1.\\
\end{array}
$$
 Moreover,
$\sum_{k=0}^{2m-1}|l_k(\t)|=[2(1+|\cos\t|)]^{m}-1$.
\end{lem}

The proof of Lemma \ref{lem3} can be found in the Appendix.
The idea is simple and direct: by viewing $l(\t)$ as the solution to an equation,
we are left to the verification when $l(\t)$ takes the form in the lemma.
Still, the computation is quite complicated and requires
special techniques from combinatorics \cite{sprugnoli_introduction_2006,sprugnoli_riordan_2007}.

\subsection{Convergence analysis and data rate}

The notations in \eqref{notation} will still be used,
except that $\psi_1$ is replaced by $\phi_1$.
The following assumptions are adopted in the subsequent analysis.

\begin{assu}\label{ass1}
The communication graph ${\mathcal{G}}$ is undirected and connected.
\end{assu}

\begin{assu}\label{ass2}
There exist known positive constants $C^\ast$ and $C_\delta^\ast$ such
that $\max\limits_{j=1,\dots,2m}||x_j(0)||\leq C^\ast$ and
$\max\limits_{j=1,\dots,2m}||\delta_j(0)||\leq
C_{\delta}^\ast$.
\end{assu}

\begin{rem}
Assumption \ref{ass1} is a standard assumption, under which the
eigenvalues of $L$ can be rearranged as
$0=\lambda_1<\lambda_2\leq\dots\leq\lambda_N$.
The reason that we only consider the undirected graph will
be clarified in Remark \ref{remdir}.
Assumption \ref{ass2} serves the same purpose as Assumption \ref{ass2m=1}.
\end{rem}

We also need the following constants:
\begin{equation}\label{constant2}
\begin{array}{rl}
   b^\ast=\max\limits_{l,j}\{|\tilde{b}_{lj}|\},&
   c^\ast=\max\limits_{1\leq j\leq 2m}\{|c_{j}|\},\\
   \Lambda_i=\max\{\lambda_i^{1/2},\lambda_i^{3/2}\},&
   \bar C=\frac{9}{\sqrt{2}}[||U_{L}^{-1}||+5c^\ast mN(||S||+2)].
\end{array}
\end{equation}

\begin{lem}\label{epsm2}
Let $\g=1-\v/4$. Then we can choose sufficiently small
$\v$ to satisfy the following inequalities:
\begin{subequations}
\begin{align}
2c^\ast\sum_{j\neq m-1}M_{ij}\v^{1/2}\le M_{i,m-1},
\;i=2,\dots,N;\label{eps3}\\
\sum_{k=0}^{2m-1}|l_k(\theta)|/{\g^{2m}}\leq [2(1+|\cos\t|)]^{m}-1/2;\label{eps2}\\
(2m-1)b^\ast(N-1)\Lambda \bar C\v^{1/2}\leq \frac{1}{8}\g^{4m-1},\label{eps4}
\end{align}
\end{subequations}
where $\Lambda=\max_{i}\Lambda_i$.
\end{lem}

\begin{thm}\label{consensus2}
Take $k_j$'s as in (\ref{kj}), $c_j$'s as in Remark
\ref{cj} and $\g=1-\v/4$. Select sufficiently small
$\v$ to satisfy Lemma \ref{epsm2} and $\rho_i<1-\v/2$ for $i=2,\dots,N$.
Then under Assumptions \ref{ass1}
and \ref{ass2}, consensus can be achieved at a convergence rate of $O(\g^t)$
provided that $M(t)$ satisfies
\begin{equation}\label{M(t)2}
  \left\{\begin{array}{l}
          M(t)\geq 1, \; t=1,\dots,2m; \\
          M(t)\geq 2^{m-1}(1+|\cos\t|)^{m}-\frac{1}{2}, ~ t=2m+1,\dots,
        \end{array}
  \right.
\end{equation}
and $p_0\geq (\sqrt{2}+1)^{2m}\max\{C^\ast,C_{\delta}^\ast\}$.

Therefore, we can use $\lceil \log_2 2\lceil
2^{m-1}(1+|\cos\t|)^{m}-\frac{1}{2}\rceil \rceil$ bits of information exchange to achieve
the consensus.
\end{thm}

\begin{IEEEproof}
1) Preparation. By (\ref{controlh}) we have
\begin{equation}\label{controllaw}
  u(t)=\left\{\begin{array}{l}
                0, ~t=0,1,\dots,2m-1;  \\
                -\sum\limits_{j=1}^{2m}k_jL(\delta_j(t)+e_j(t)),~ t\geq 2m.
              \end{array}
  \right.
\end{equation}

Direct computation shows
$$
\left[\begin{array}{c}
          \delta_{1}(t+1) \\
          \vdots \\
          \delta_{2m}(t+1)
        \end{array}\right]=(A\otimes I_N)\left[\begin{array}{c}
          \delta_{1}(t) \\
          \vdots \\
          \delta_{2m}(t)
        \end{array}\right]+\left[\begin{array}{c}
          0 \\
          \vdots \\
          u(t)
        \end{array}\right]
$$
Let $\tilde{\delta}_j(t)
=U_{L}^{-1}{\delta}_j(t)=
[\tilde{\delta}_{1j}(t),\dots,\tilde{\delta}_{N,j}(t)]^T$
 and
$\tilde{\delta}^{i}(t)=[\tilde{\delta}_{i1}(t),\dots,\tilde{\delta}_{i,2m}(t)]^T$.
Then we obtain $\tilde{\delta}^{1}(t)\equiv 0$, and
for $i=2,\dots,N$
\begin{equation}\label{distilde}
\tilde{\delta}^{i}(t+1)=\left\{
                          \begin{array}{ll}
                            A\tilde{\delta}^{i}(t), & t=0,1\dots,2m-1; \\
                            A_i\tilde{\delta}^{i}(t)-\epsilon_i(t), & t\geq 2m,
                          \end{array}
                        \right.
\end{equation}
where $\epsilon_i(t)=[0,\dots,0,\sum_{j=1}^{2m}k_j\lambda_i\phi_{i}^{T}e_{j}(t)]^T\in \R^{2m}$.

2) Estimation error and exponential convergence. 
To analyze the influence of $u(t)$ on the error term $e_j(t)$,
we will show $|\phi_{i}^{T}u(t)|\leq
\Lambda_i \bar Cp_0\g^{t-2m}\v^{1/2},t\geq 2m$ by induction.

With the choice of $p_0$ and $\g$ it's easy to see
$|s_i(t)|\leq 3/2$ when $t\leq 2m$ by noticing
$|y_i(t)|\leq (\sqrt{2}+1)^{2m}C^\ast$, hence we obtain
$\max\limits_{1\leq t\leq 2m}||\Delta(t)||\leq 1/2$
provided $M(t)\geq1,t=1,\dots,2m$. Moreover,
$||\delta_j(2m)||\leq (\sqrt{2}+1)^{2m}C_{\delta}^\ast$.
Recalling (\ref{controlh}) and $e_j(2m)\leq p_0||S||
\max\limits_{1\leq t\leq 2m}||\Delta(t)||$
we have
\begin{equation*}
\begin{aligned}
|\phi_{i}^{T}u(2m)|& = |\phi_{i}^{T}\sum_{j=1}^{2m}k_jL(\delta_j(2m)+e_j(2m))| \\
&\le 2\lambda_imNc^*\v((\sqrt{2}+1)^{2m}C_{\delta}^\ast+p_0||S||/2)\\
&\le 2\lambda_imNc^*\v(1+||S||/2)p_0\\
&\le \Lambda_i\bar Cp_0\v^{1/2}
\end{aligned}
\end{equation*}
by noticing $p_0\ge (\sqrt{2}+1)^{2m}C_{\delta}^\ast$ and $\v<1$.

Assume that
\begin{equation}\label{assinduct}
\begin{aligned}
|\phi_{i}^{T}u(\tau)| &\leq \Lambda_i \bar C p_0\g^{\tau-2m}\v^{1/2},\;2m \le \tau\leq t; \\
||\Delta(\tau)|| &\leq \frac{1}{2} (\Rightarrow |e_{i1}(\tau)|\le \frac{1}{2}p_0\g^{\tau-1}),\; 1\leq \tau \leq t.
\end{aligned}
\end{equation}
$$
\begin{array}{l}
  e_j(\tau)=\left[
\begin{array}{l}
e_{11}(\tau-2m+1) ~ \dots ~ e_{11}(\tau) \\
\quad\quad\quad\vdots \quad\quad\quad\quad \ddots \quad \vdots \\
e_{N1}(\tau-2m+1) ~ \dots ~ e_{N1}(\tau) \\
\end{array}
\right]\left[
\begin{array}{c}
S(j,1) \\
\vdots \\
S(j,2m) \\
\end{array}
\right] -\sum_{n=1}^{2m-1}\tilde{b}_{nj}(\t)u(\tau-n),
\end{array}
$$
we have
\begin{equation*}
\begin{aligned}
|\phi_{i}^{T}e_j(\tau)|&\le Np_0\max\limits_{\tau-2m+1\leq s\leq
	\tau}||\Delta(s)||\g^{\tau-2m}||S||  +N(2m-1)b^\ast \lambda_i\v^{1/2}\bar Cp_0\g^{\tau-4m+1}
\end{aligned}
\end{equation*}
and
\begin{equation}\label{Ei}
\begin{array}{rl}
  ||\epsilon_i(\tau)||
  &\le \lambda_ic^\ast\v2mN[p_0\max\limits_{\tau-2m+1\leq s\leq \tau}||\Delta(s)||\g^{\tau-2m}||S||+(2m-1)b^\ast \lambda_i\v^{1/2} \bar Cp_0\g^{\tau-4m+1}]\\
&\le 2\lambda_ic^\ast mN(||S||/2+1)p_0\g^{\tau-2m}\v
\end{array}
\end{equation}
by \eqref{eps4},
if $||\Delta(s)||\leq 1/2$ for $\tau-2m+1\leq s\leq \tau$.
Recalling (\ref{distilde}) we obtain
$$
\tilde{\delta}^{i}(t+1)=A_{i}^{t+1-2m}\tilde{\delta}^{i}(2m)
-\sum_{\tau=0}^{t-2m}A_{i}^{t-2m-\tau}\epsilon_i(2m+\tau).
$$
By applying Lemma \ref{lem2} and taking into account (\ref{Ei}), it yields that
\begin{equation}\label{diff_esti}
\begin{array}{l}
   \quad |\tilde{\delta}_{i,2j-1}(t+1)|,|\tilde{\delta}_{i,2j}(t+1)| \\
    \leq  \left\{\begin{array}{l}
   M_{ij}\v^{j-(m-1)}\g^{t+1-2m}
   \big[||\tilde{\delta}^{i}(2m)||+
   4\lambda_ic^*mN (||S||+2)p_0\big], \\
    j=1,\dots,m-2; \\
    M_{ij}\v^{(j-m)/2}\g^{t+1-2m}
    \big[||\tilde{\delta}^{i}(2m)||+
   4\lambda_ic^*mN (||S||+2)p_0\big], \\
    j=m-1,m,
  \end{array}\right.
\end{array}
\end{equation}
due to $\v/(\g-\rho_i)<4$.

With \eqref{diff_esti} it is ready to estimate $\phi_{i}^{T}u(t+1)$,
which is a sum of $\sum_{j=1}^{2m}k_j\phi_{i}^{T}L{\delta}_{j}(t+1)$
and $\sum_{j=1}^{2m}k_j\phi_{i}^{T}e_j(t+1)$.
For the first part, by \eqref{diff_esti} and
$||\tilde{\delta}^{i}(2m)||\le ||U_{L}^{-1}||||{\delta}_i(2m)||$
we have
\begin{equation}\label{part1}
\begin{aligned}
   & \quad|\sum_{j=1}^{2m}k_j\phi_{i}^{T}L{\delta}_{j}(t+1)| \\
  & = \lambda_i|\sum_{j=1}^{2m}k_j\tilde{\delta}_{ij}(t+1)|\\
  & \le \lambda_ip_0\g^{t-2m+1}(2M_{i,m-1}\v^{1/2}+2c^\ast\v
  \sum_{j\neq m-1}M_{ij})\cdot\\
&\quad[||\tilde{\delta}^{i}(2m)||+4\lambda_ic^*mN(||S||+2)p_0]\\
  &\le 3\lambda_iM_{i,m-1}[||U_{L}^{-1}||+4\lambda_ic^*mN(||S||+2)]p_0\g^{t-2m+1}\v^{1/2}\\
\end{aligned}
\end{equation}
if we note that $|c_{2m-2}|,|c_{2m-3}|\le 1$.
For the second part, as in the second order case,
it is closely related with $||\Delta(t+1)||$
and similarly it can be inferred from \eqref{di(t)m2} that
$d(t+1)$ is only dependent on the past
quantization errors $\Delta(\tau), t-2m+1\leq \tau\leq t$
and the past control inputs $u(\tau), t-2m+1\leq \tau\leq t-1$.
Hence with the induction assumption (\ref{assinduct}) the quantizer can be made
unsaturated at time $t+1$ with finite bits, namely $||\Delta(t+1)||\leq 1/2$.
In consequence we get an estimation similar to \eqref{Ei} that
\begin{equation}\label{part2}
\begin{array}{rl}
    |\sum_{j=1}^{2m}k_j\phi_{i}^{T}Le_{j}(t+1)|
  \le 2\lambda_ic^\ast mN(||S||/2+1)p_0\g^{t-2m+1}\v.
\end{array}
\end{equation}
Combining \eqref{part1} and \eqref{part2}, it is clear that
$
|\phi_{i}^{T}u(t+1)|\le \Lambda_i \bar C p_0\g^{t-2m+1}\v^{1/2},
$
which establishes the induction.
Furthermore, by (\ref{diff_esti})
clearly the consensus can be achieved at a
convergence rate of $O(\g^t)$.

3) Data rate. Below we are to discuss the number of quantization levels at each time
step. The situation when $t\leq 2m$ has been discussed.
When $t> 2m$, we have
$$
\begin{array}{l}
  \quad||d(t)||\\
  \leq  ||\frac{\cos\t}{\g}\Delta(t-1)+
  \sum\limits_{j=1}^{2m}\frac{\sin\t S(2,j)+S(3,j)}{\g^{2m-j+1}}\Delta(t-1-2m+j)||
+2b^\ast\sum_{j=1}^{2m-1}\frac{||u(t-1-j)||}{p(t-1)}\\
  =||\sum\limits_{j=1}^{2m} \frac{1}{\g^{2m-j+1}}[\cos\t S(1,j)+ \sin\t S(2,j)+S(3,j)]\Delta(t-1-2m+j)||
  \\\quad+2b^\ast\sum_{j=1}^{2m-1}\frac{||u(t-1-j)||}{p(t-1)}\\
  \leq \frac{1}{2\g^{2m}}\sum_{k=0}^{2m-1}|l_k(\theta)|+
  2b^\ast(2m-1)(N-1)\Lambda \bar C\g^{1-4m}\v^{1/2}
\end{array}
$$
by noticing $S(1,\cdot)=[0,\dots,0,1]
$ and $u(t)=\sum_{i=1}^{N}\phi_{i}\phi_{i}^{T}u(t)$,
$\Lambda_1=0$.
By taking into account \eqref{eps2} and \eqref{eps4} it can be seen that
$||d(t)||$ is bounded by $2^{m-1}(1+|\cos\t|)^{m}$ and
the proof is completed by remembering (\ref{M(t)2}).
\qquad\end{IEEEproof}

\begin{rem}
Noticing that $\prod_{1\leq k\leq m-2,k\neq n}|k-n|=(n-1)!(m-2-n)!$ attains the minimum at
$n=\lfloor \frac{m-2}{2}\rfloor$ and multiplying by a positive $\lambda_i$ on both sides
does not change the direction of an inequality,
\eqref{eps3} can be substituted by the following stronger one,
which is easier to check:
\begin{equation}\label{ineq1}
\begin{array}{c}
  5c^*(\sum_{j=1}^{m-2}\sum_{n=1}^{m-2}\frac{n^{j-1}}
  {(\lfloor \frac{m}{2}\rfloor-2)!(m-1-\lfloor \frac{m}{2}\rfloor)!}
  +1+\lambda_N)\v^{1/2}  \leq 3\sqrt{\frac{\lambda_2}{2}}.
\end{array}
\end{equation}
\end{rem}

\begin{rem}\label{remdir}
From the proof it is readily seen that we can still use the same number of bits
to achieve the quantized consensus once the
Laplacian of the directed topology satisfies that $0<\l_2\le\dots\le\l_N$.
However, unlike the case of the 2nd-order oscillator, it does not hold for the general topology,
when the Laplacian contains complex eigenvalues, or real Jordan blocks of multiple dimensions.
For one reason, note that Lemma \ref{lem1} does not hold for a complex $\l_i$.
For another one, note the disparity in the order of $\v$
between the disturbance term and the weighted sum of disagreement entries, i.e.
$||\epsilon_i(t)||\sim O(\v)p_0\g^t$ and $||K\tilde\delta^i(t)||\sim O(\v^{1/2})p_0\g^t$.
Therefore, if we assume $m=2$ and the Jordan block corresponding to $\l_2>0$ 
is two-dimensional as in \eqref{disagreement}, 
then it follows from $||K\tilde\delta^3(t)||\sim O(\v^{1/2})p_0\g^t$
that $||\tilde\delta^2(t)||\sim O(\v^{-1})p_0\g^t$
and $||u(t)||\sim O(1)p_0\g^t$, suggesting that the input term can no longer be neglected in the estimation errors,
nor in the quantization input $d(t)$. Such a situation is also encountered in \cite{qiu_quantized_2016}.
\end{rem}

\begin{rem}
At the first glance it may seem doubtful that the data rate is dependent on $|\cos\t|$;
but a little further inspection is enough to clarify.
Similar to the situation of the $n$-th order integrator system investigated in \cite{qiu_quantized_2016},
the control input does not consume any bit in exchanging the estimates of the states when $\v$ is sufficiently small.
In other words, we only need to focus on how many bits it needs to estimate the output
of an individual open loop system. Take the second-order case as an example.
Noticing that $y_i(t)=\cos\t{x}_{i1}(t-1)+\sin\t{x}_{i2}(t-1)=2\cos\t{x}_{i1}(t-1)-{x}_{i1}(t-2)$,
we can estimate $y_i(t)$ based on $\hat{x}_{i1}(t-1)$ and $\hat{x}_{i1}(t-2)$
with an error bound no larger than $\frac{1}{2}(2|\cos\t|+1)+\frac{1}{2}$.
Generally speaking, when $|\cos\t|\approx 0$ or equivalently $|\sin\t|\approx 1$,
$x_{i,2j-1}(t)$ and $x_{i,2j}(t)$ are tightly coupled,
and it needs only $m$ bits of information exchange to achieve the consensus;
in the case of $|\cos\t|\approx 1$, after rearranging of states $A$ can be approximated by
$I_2\otimes J_{1,m}$, and $2m$ bits are sufficient.
Anyway, for a $2m$-th order system studied in this paper,
$2m$ bits are enough to realize the consensus asymptotically,
which is consistent with the conclusion for $n$-th order integrator systems \cite{qiu_quantized_2016}.
\end{rem}

\section{Numerical Example}\label{sec:example}

For simplicity we only show an example of $m=2$.
Consider a 5-node network with 4-th order dynamics,
where the edges are generated randomly according to
probability $P(i,j)\in \mathcal{E}=0.5$ with 0-1 weights.
The initial states are randomly chosen as $x_{ij}(0)\in (0,j)$, $i=1,\dots,5,j=1,\dots,4$.
Given $\t=\pi/3$, it is enough to use 3 bits of information exchange to realize the consensus,
and we can compute $S_2=\left[
                          \begin{array}{cccc}
                             -4/(3\sqrt 3) & 2/\sqrt 3  & -4/(3\sqrt 3) &  5/(3\sqrt 3) \\
                            -1/3  &  1  & -1  &  2/3 \\
                            -1/\sqrt 3  &  1/\sqrt 3  & -1/\sqrt 3  &  0\\
                          \end{array}
                        \right]
$ to construct the encoder and decoder respectively as \eqref{encoder1}-\eqref{decoder2}.
The communication topology is generated as in Figure 1
with $\lambda_2=0.8299$, and $c_j$'s are determined as in Remark \ref{cj} by choosing $h=\l_2$.
Moreover, let $\v=0.01$, $p_0=10$, $\g=0.9975$ to satisfy the conditions in Theorem \ref{consensus2}.
From Figure 2 which depicts the trajectory of
$\delta_{jmax}(t)=\{\delta_{nj}(t):\,n=\text{arg}
\max_{i}|\delta_{ij}(t)|\}$,
we can see that the consensus is achieved asymptotically.

\begin{figure}
  \center
  \includegraphics[width=0.7\linewidth]{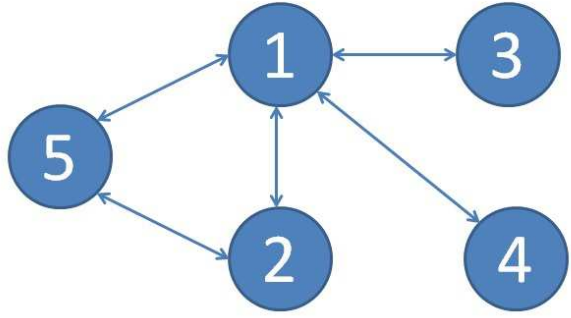}
  \caption{Communication topology}
\end{figure}

\begin{figure}
\begin{center}$
\begin{array}{cc}
  \includegraphics[width=0.45\linewidth]{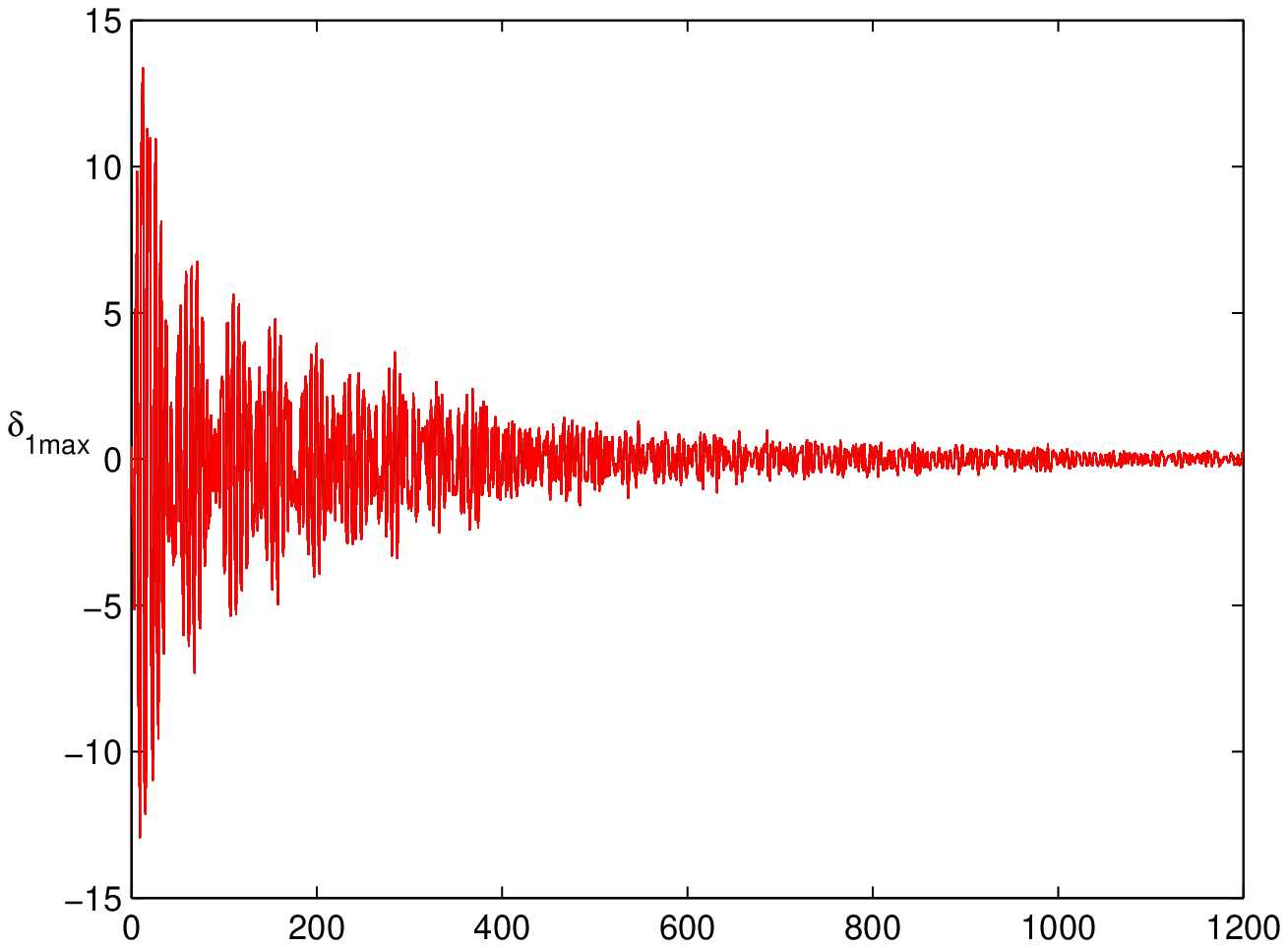} &
  \includegraphics[width=0.45\linewidth]{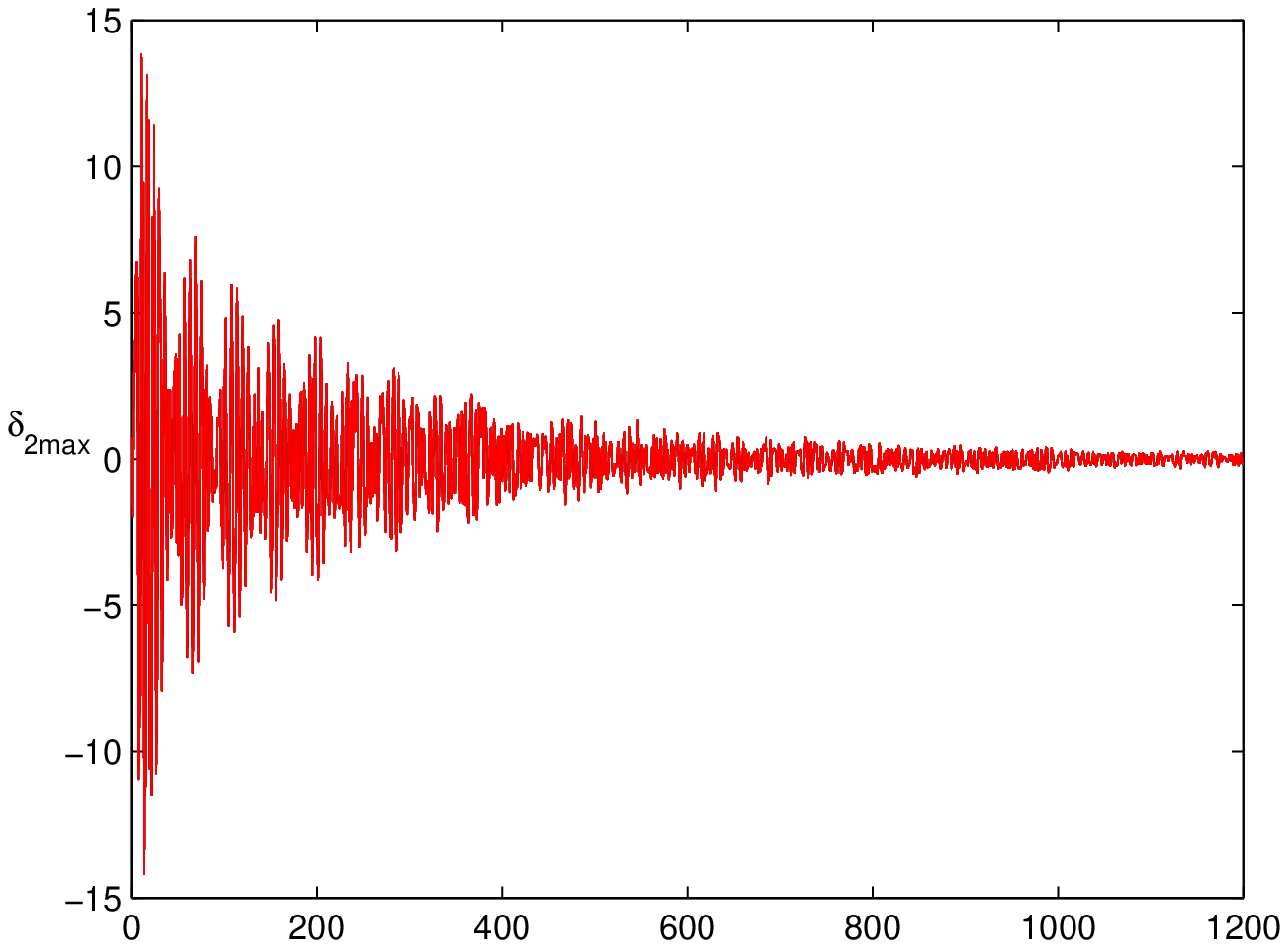} \\
  \includegraphics[width=0.45\linewidth]{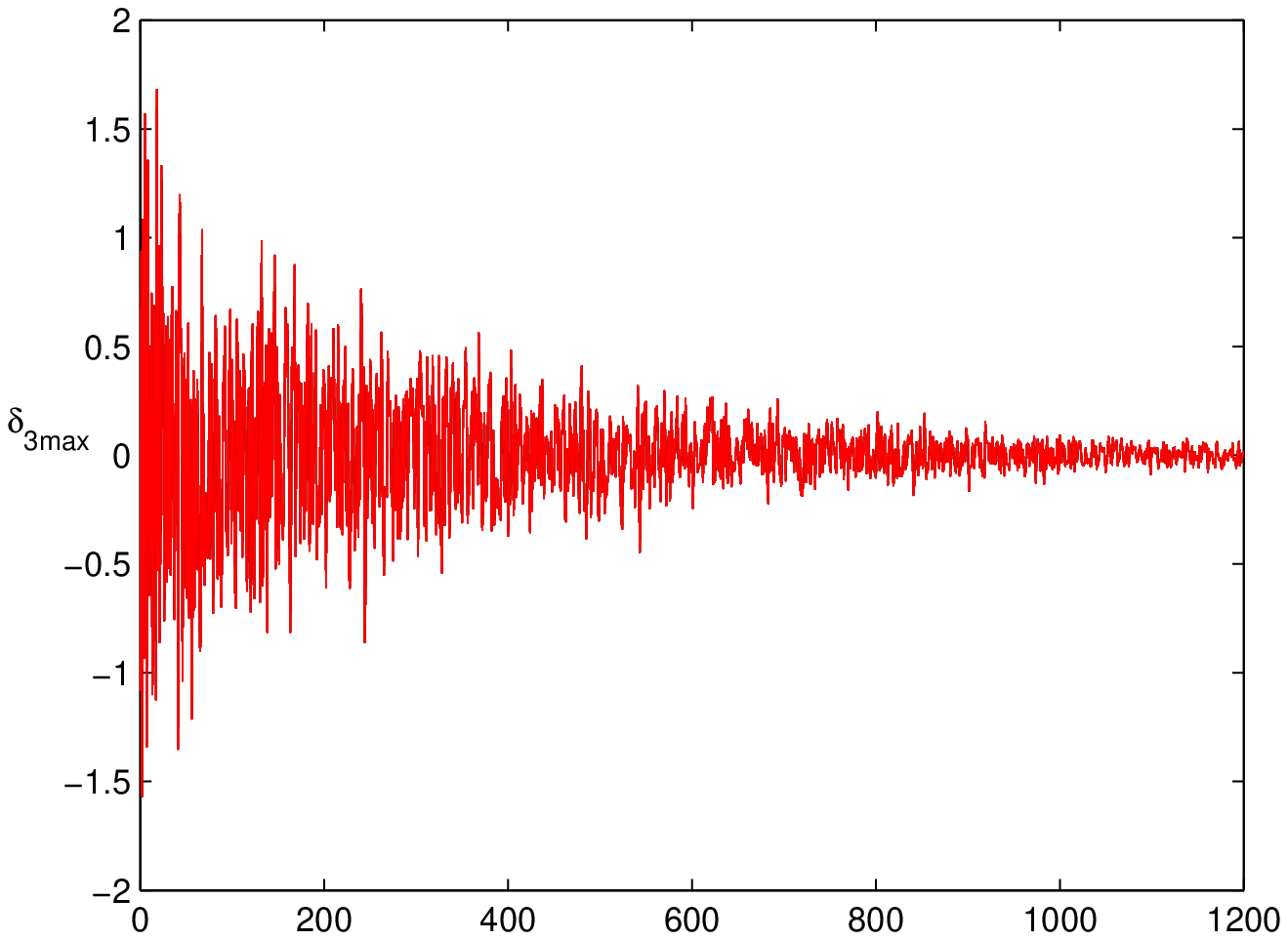} &
  \includegraphics[width=0.45\linewidth]{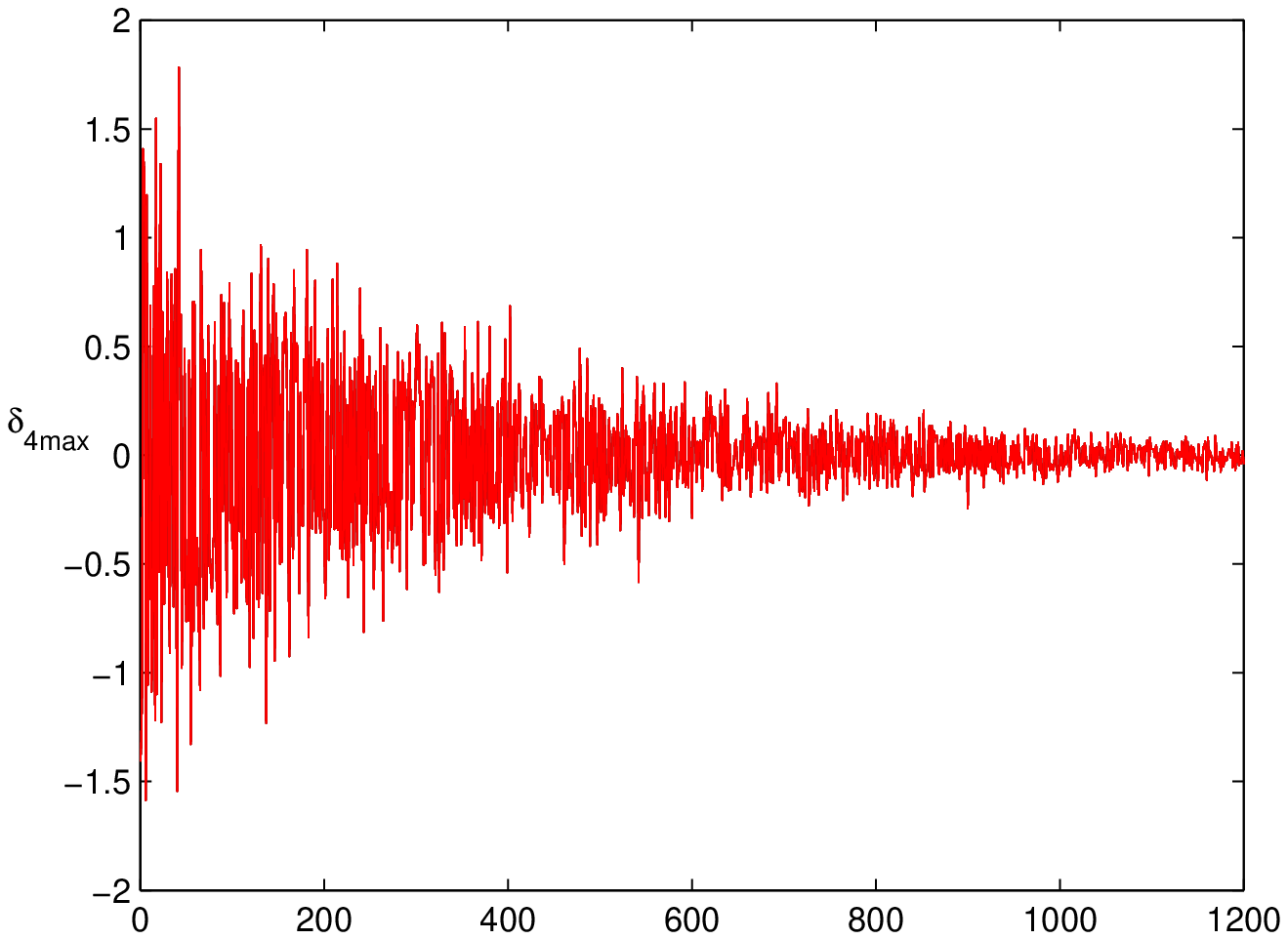}
\end{array}$
\end{center}
  \caption{Trajectories of disagreements $\delta_{jmax}(t)$}
\end{figure}

\section{Concluding Remarks}\label{sec:conclusion}

In this paper, we explored the data rate problem for quantized consensus of
a special kind of multi-agent systems. The dynamics of each agent
is described by a $2m$-th order real Jordan form
consisting of $m$ pairs of conjugate poles on the unit circle with single input,
and only the first state can be measured.
The encoding-decoding scheme was based on the observability matrix.
Perturbation techniques were employed in the consensus analysis and the data rate analysis,
and combinatorial techniques were used to explicitly obtain the data rate.
The second-order case of $m=1$ and higher-order cases of $m\ge 2$ were investigated separately.
For the second-order case, we showed that at most 2 bits of information exchange
suffice to achieve the consensus at an exponential rate,
if the communication topology has a spanning tree.
For the higher-order cases, consensus was achieved with at most $2m$ bits,
provided that the undirected communication topology is connected.
The exact number of bits for achieving consensus in both cases
is an integer which increases from $m$ to $2m$ when $|\cos\t|$ increases from 0 to 1.
The case of switching directed topology is still under
investigation, and noisy communication channels will be considered in the future work.
As for general unstable systems with poles outside the unit circle,
perturbation techniques no longer apply and
new methods need to be developed to serve the same purpose of stabilizing
the dynamics of disagreements.

\appendix
{\em Proof of Lemma \ref{lem1}}
Here we mainly deal with the case of $m\ge 3$, since the proof can be slightly adapted
if $m=2$ and the modification will be pointed out accordingly.
The characteristic equation of $A_i$ can be computed as
\begin{align*}
\chi_i(\mu)=\det[(\mu I-Q)^m+\lambda_iK_m(\mu I-Q)^{m-1}+\dots+\lambda_iK_2(\mu I-Q)+\lambda_iK_1],
\end{align*}
where $Q=\left[
          \begin{array}{cc}
            \cos \t & \sin \t \\
            -\sin \t & \cos \t \\
          \end{array}
        \right]$ and
$K_j=\left[
          \begin{array}{cc}
            0 & 0 \\
            k_{2j-1} & k_{2j} \\
          \end{array}
        \right]$ for $j=1,\dots,m$.
By employing \eqref{P} in the proof of Lemma \ref{lemm=1}, we rewrite $\chi_i(\mu)$ as
\begin{equation}\label{character}
\begin{aligned}
  \chi_i(\mu)& = (\mu-e^{\jmath\t})^m(\mu-e^{-\jmath\t})^m
  +\frac{1}{2}[(\mu-e^{\jmath\t})^m+(\mu-e^{-\jmath\t})^m]\frac{\lambda_i}{2}\cdot \\
   &\quad [\sum\limits_{j=2}^{m}(-k_{2j-1}\jmath+k_{2j})(\mu-e^{\jmath\t})^{j-1}+2k_2
   +\sum\limits_{j=2}^{m}(k_{2j-1}\jmath+k_{2j})(\mu-e^{-\jmath\t})^{j-1}]\\
   &\quad +\frac{\jmath}{2}[(\mu-e^{\jmath\t})^m-(\mu-e^{-\jmath\t})^m]\frac{\lambda_i}{2}\cdot \\
   &\quad [\sum\limits_{j=2}^{m}(k_{2j-1}+k_{2j}\jmath)(\mu-e^{\jmath\t})^{j-1}+2k_1
   +\sum\limits_{j=2}^{m}(k_{2j-1}-k_{2j}\jmath)(\mu-e^{-\jmath\t})^{j-1}]\\
   &= (\mu-e^{\jmath\t})^m[(\mu-e^{-\jmath\t})^m+\frac{\lambda_i}{2}(k_2+k_1\jmath)] +\frac{\lambda_i}{2}(\mu-e^{-\jmath\t})^m\sum\limits_{j=1}^{m}(\mu-e^{\jmath\t})^{j-1}(-k_{2j-1}\jmath+k_{2j}).
\end{aligned}
\end{equation}
With $A_i$ being real, we only need to focus on the perturbed roots around $e^{\jmath\t}$,
which are denoted by $\mu=e^{\jmath\t}+\Delta\mu$.
Noticing that $\mu-e^{-\jmath\t}=\mu-e^{\jmath\t}+2\sin\t\jmath$,
we substitute $\mu=e^{\jmath\t}+\Delta\mu$ into \eqref{character} and obtain
\begin{equation}\label{chi2}
  \chi_i(e^{\jmath\t}+\Delta\mu) = \sum_{n=1}^{m}a_{in}(\v)(\Delta\mu)^{m-n}+\sum_{n=0}^m (C(m,m-n)(2\sin\t\jmath)^{m-n}+O(\v))(\Delta\mu)^{m+n}
\end{equation}
with the selection of $k_{2j-1}$ and $k_j$ in \eqref{kj},
where
$$
a_{in}(\v)=\left\{
             \begin{array}{l}
               (2\sin\t\jmath)^{m-1}[2\sin\t\jmath(c_{2m}-c_{2m-1}\jmath)
+m(c_{2m-2}-c_{2m-3}\jmath)]\v+o(\v),
 ~ n=1; \\
             (2\sin\t\jmath)^{m}(c_{2(m-n+1)}-c_{2(m-n)+1}\jmath)\v^{n-1}+ o(\v^{n-1}) , ~ n=2,\dots,m.
             \end{array}
           \right.
$$
\begin{figure}
  \center
  \includegraphics[height=4cm,width=9.5cm]{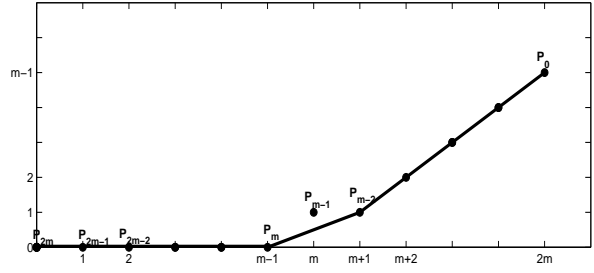}
  \caption{Newton diagram}
\end{figure}
Now the Newton diagram \cite{seyranian_multiparameter_2003} can be depicted as in Fig. 3,
by first plotting points $P_{2m-j}(j,\alpha_{2m-j})$, $j=0,\dots,2m$ and
then connecting the segments on the lower boundary of the convex hull of the above points,
where $\alpha_{2m-j}$ is the leading exponent of $\v$ in the coefficient of $(\Delta\mu)^{2m-j}$.
The slopes of the two non-horizontal segments are 1/2, 1 respectively, implying that
$\Delta\mu$ has the following two forms of expansions:
\begin{subequations}
\begin{align}
\Delta\mu  =\mu_1\v^{\frac{1}{2}}+\mu_2\v^{\beta}+o(\v^{\beta}),~\mu_1\ne 0;\label{mu1}\\
\Delta\mu  =\upsilon_1\v+o(\v),~\vartheta_1\ne 0.\label{mu2}
\end{align}
\end{subequations}

Substituting (\ref{mu1})
into (\ref{chi2}) and finding the coefficients of the term $\v^{m/2}$,
it yields that
$\mu_{1}^{m}(2\jmath\sin\t)^m+\mu_{1}^{m-2}\frac{\lambda_i}{2}(2\jmath\sin\t)^m(c_{2m-2}-c_{2m-3}\jmath)=0$,
and thus
\begin{equation}
\mu_{1}=\pm\sqrt{\frac{\lambda_i}{2}\sqrt{c_{2m-3}^{2}+c_{2m-2}^{2}}}e^{\jmath\frac{\alpha}{2}},
\end{equation}
where $\alpha=\arg(c_{2m-3}\jmath-c_{2m-2})$.
Moreover, to determine $\mu_2$ and $\beta$,
we substitute (\ref{mu1}) into (\ref{chi2}) again and find the lowest order term as
\begin{equation}
\begin{array}{l}
    \quad m\mu_{1}^{m-1}\mu_{2}(2\jmath\sin\t)^m\v^{(m-1)/2}\v^{\beta}\\
    +\frac{\lambda_i}{2}\mu_{1}^{m-1}(2\jmath\sin\t)^m(c_{2m}-c_{2m-1}\jmath)\v^{(m+1)/2} \\
  + \frac{\lambda_i}{2}(m-2)\mu_{1}^{m-3}\mu_{2}(2\jmath\sin\t)^m(c_{2m-2}-c_{2m-3}\jmath)\v^{(m-1)/2}\v^{\beta}\\
    +\frac{\lambda_i}{2}\mu_{1}^{m-3}(2\jmath\sin\t)^m(c_{2m-4}-c_{2m-5}\jmath)\v^{(m+1)/2}=0,
\end{array}
\end{equation}
which implies $\beta=1$ and
$\mu_{2}=\frac{\lambda_i}{4}(c_{2m-1}\jmath-c_{2m})
+\frac{(c_{2m-5}\jmath-c_{2m-4})}{2(c_{2m-3}\jmath-c_{2m-2})}$.
In the form of (\ref{mu1}), the module of $\mu$ is determined as
\begin{equation}
\begin{array}{l}
  |\mu|^2=  \mu\bar{\mu}
  =  1+2\text{Re}(\mu_1e^{-\jmath\t})\v^{\frac{1}{2}}
  +(|\mu_1|^2+2\text{Re}(\mu_2e^{-\jmath\t}))\v+o(\v),
\end{array}
\end{equation}
with
$\text{Re}(\mu_1e^{-\jmath\t})=\pm\sqrt{\frac{\lambda_i}{2}\sqrt{c_{2m-3}^{2}+c_{2m-2}^{2}}}\cos(\frac{\alpha}{2}-\t)$.
In order that $|\mu|<1$ with sufficiently small $\v$, we must have $\frac{\alpha}{2}-\t=\frac{\pi}{2}$,
and hence it suffices to let $|\mu_1|^2+2\text{Re}(\mu_2e^{-\jmath\t})<0$.
Combining these arguments gives rise to a sufficient condition as
\begin{subequations}
\begin{gather}
-c_{2m-3}/c_{2m-2}=\tan(2\t+\pi), \; c_{2m-3}^{2}+c_{2m-2}^{2}\neq 0;\label{con1a}\\
  \frac{\lambda_i}{2}\sqrt{c_{2m-3}^{2}+c_{2m-2}^{2}}+\frac{\lambda_i}{2}(c_{2m-1}\sin\t-c_{2m}\cos\t)
  +\text{Re}[\frac{c_{2m-5}\jmath-c_{2m-4}}{c_{2m-3}\jmath-c_{2m-2}}e^{-\jmath\t}]<0.\label{con1b}
\end{gather}
\end{subequations}
With $c_{2m-3}=-\sin2\t$ and $c_{2m-2}=\cos2\t$ satisfying \eqref{con1a},
\eqref{con1b} is equivalent to $\lambda_iR_m+H<0$.
When $m=2$, $\mu$ only takes the form of \eqref{mu1} and
$\mu_{2}=\frac{\lambda_i}{4}(c_{2m-1}\jmath-c_{2m})$,
leading to the sufficient condition $R_2<0$ for $|\mu|<1$.

On the other hand, substituting (\ref{mu2})
into (\ref{character}) and finding the coefficients of the term $\v^{m-1}$,
we obtain the equation \eqref{eqn}.
Similarly, the module of $\mu$ with the form (\ref{mu2}) is determined by
\begin{equation}
  |\mu|^2=1+2\text{Re}(\vartheta_1e^{-\jmath\t})\v+o(\v)
\end{equation}
and it suffices to let $\text{Re}(\vartheta_1e^{-\jmath\t})$ to be negative such that
$|\mu|<1$ with sufficiently small $\v$.
For prescribed $c_{2m-3}$ and $c_{2m-2}$, the roots of (\ref{eqn})
can be assigned arbitrarily such that $\text{Re}(\vartheta_1e^{-\jmath\t})<0$ with $m-2$ distinct $\vartheta_1$;
after determining $c_{2m-4}$ and $c_{2m-5}$, (\ref{con1b}) can always be satisfied
by properly chosen $c_{2m}$ and $c_{2m-1}$ since $\frac{\lambda_i}{2}(c_{2m-1}\sin\t-c_{2m}\cos\t)$
can be assigned to any number. In summary, the proof is completed. \IEEEQEDclosed

{\em Proof of Lemma \ref{lem2}}
As in the last proof, we only focus on the case of $m\ge 3$
which essentially includes the case of $m=2$.
For $A_i$, we are to find the following Jordan decomposition:
\begin{equation}\label{Aij}
A_{i}= A+\sum_{j=1}^{m-1}A_{ij}\v^j=R_i\tilde{A}_iR_{i}^{-1},
\end{equation}
where $\tilde{A}_i$ is a diagonal matrix consisting of $2m$ different
eigenvalues determined in Lemma \ref{lem1}.
To find an appropriate $R_i$ and the corresponding $R_{i}^{-1}$,
we first determine the Jordan basis of the unperturbed matrix $A$.
The Jordan chain corresponding to the eigenvalue $\mu_0=e^{\jmath\t}$ is given by
$$
u_{m-1} \xrightarrow{A-\mu_0I} u_{m-2} \xrightarrow{A-\mu_0I}\dots \xrightarrow{A-\mu_0I}u_1 \xrightarrow{A-\mu_0I} u_0,
$$
where $u_j=e_{2j+1}+\jmath e_{2j+2},j=0,\dots,m-1$ and $e_n\in \R^{2m}$
denotes the vector with a 1 in the $n$-th coordinate and 0's elsewhere.
Similarly, the Jordan chain corresponding to the eigenvalue $\bar{\mu}_0=e^{-\jmath\t}$ is given by
$$
\bar{u}_{m-1} \xrightarrow{A-\bar{\mu}_0I} \bar{u}_{m-2}
\xrightarrow{A-\bar{\mu}_0I}\dots \xrightarrow{A-\bar{\mu}_0I}\bar{u}_1 \xrightarrow{A-\bar{\mu}_0I} \bar{u}_0.
$$
Hence the two Jordan chains of $A$ can be rearranged as
$R_0=(u_0~\bar{u}_0~\dots~u_{m-1}~\bar{u}_{m-1})=I_m\otimes P$ with
                               $P=\left[
                                   \begin{array}{cc}
                                     1 & 1 \\
                                     \jmath & -\jmath \\
                                   \end{array}
                                 \right]$.
With $A_i$ being real, once we obtain the eigenvectors corresponding to
the $m$ different perturbed eigenvalues around $\mu_0$, the other eigenvectors can be obtained
by taking conjugates. Hence we only need to find the eigenvectors corresponding to
the $m$ different perturbed eigenvalues around $\mu_0$.

The eigenvectors corresponding to the $m$ perturbed eigenvalues around $\mu_0$
have the following form of Puiseux series \cite{baumgartel_analytic_1985}:
$$
\begin{array}{l}
  \mu_{in}=\mu_0+\sum_{k=1}^{\infty}\mu_{ink}\v^{k/2},
  ~ u_{in}=u_{in0}+\sum_{k=1}^{\infty}u_{ink}\v^{k/2},~n=1,2; \\
  \mu_{in}=\mu_0+\sum_{k=1}^{\infty}\vartheta_{ink}\v^{k},
  ~ u_{in}=u_{in0}+\sum_{k=1}^{\infty}u_{ink}\v^{k},~n=3,\dots,m,
\end{array}
$$
where $\mu_{i11}=\sqrt{\frac{\lambda_i}{2}\sqrt{c_{2m-3}^{2}+c_{2m-2}^{2}}}e^{\jmath\frac{\alpha}{2}}$,
$\mu_{i21}=-\mu_{i11}$ and
$\vartheta_{in1}=\vartheta_{n1},n=3,\dots,m$ have been defined in Lemma \ref{lem1}.
Substituting $\mu_{in}, u_{in}$ into the equation $A_iu_{in}=\mu_{in} u_{in}$ respectively,
and collecting coefficients of equal powers of $\v$;
moreover, noticing the fact that $A_{ij}u_k=0,j=1,\dots,m-2;k=0,\dots,m-2-j,$
where $A_{ij}$ has been defined in \eqref{Aij}
and imposing the normalization condition as $v_{m-1}^{T}u_{in}=1$,
where $v_{m-1}^{T}=\frac{1}{2}\left[
                                \begin{array}{ccccc}
                                  1 & -\jmath & 0 & \dots & 0 \\
                                \end{array}
                              \right]
$ is the left associated eigenvector of $A$ with respect to the eigenvalue $\mu_0$
such that $v_{m-1}^{T}u_0=1,v_{m-1}^{T}u_1=\dots=v_{m-1}^{T}u_{m-1}=0$,
the $m$ eigenvectors can be obtained as:
$$
\begin{array}{c}
   u_{in}=u_0+\sum\limits_{k=1}^{m-1}\v^{k/2}(\mu_{i1n}^{k}u_k+u_{ikn}^{'})+o(\v^{m/2}),~n=1,2;  \\
   u_{in}=u_0+\sum\limits_{k=1}^{m-1}\v^{k}(\vartheta_{1n}^{k}u_k+u_{ikn}^{'})+o(\v^{m}),~n=3,\dots,m,
\end{array}
$$
where
$u_{i1n}^{'}=0$ and
$u_{ikn}^{'}\in \text{span}\{u_1,\dots,u_{k-1}\},k=2,\dots,m-1$ for $n=1,\dots,m$.

Letting $R_i=\left[
           \begin{array}{ccccc}
             u_{i1} & \bar u_{i1} & \dots & u_{im} & \bar u_{im} \\
           \end{array}
         \right]
$, we are to investigate the magnitude of each entry in $R_{i}^{-1}$ by adjoint method.
Therefore we need to find the order of $\det R_{i}$ and the corresponding cofactor,
both of which can be expressed as Puiseux series.
The following facts should be mentioned before the calculation:

\noindent 1). Determinant is a multi-linear function of column vectors,
and it vanishes when two or more columns coincide.

\noindent 2). There exist two types of series in the columns of $R_i$,
and we categorize $u_{i1},u_{i2}$ and their conjugates for type I,
the others for type II.

With these facts, we can see that the lowest degree can be obtained by taking out
terms with $\v^{(m-2)/2}u_{m-2}$ and $\v^{(m-1)/2}u_{m-1}$ respectively from $u_{i1}$ and $u_{i1}$,
terms with $u_0,\v u_1,\dots,\v^{m-3} u_{m-3}$ respectively from $u_{i3},\dots,u_{im}$,
as well as the corresponding conjugates from $\bar u_{i1},\dots,\bar u_{im}$,
and calculated by $2(0+1+\dots+m-3+\frac{m-2}{2}+\frac{m-1}{2})=m^2-3m+3$.
Moreover,
\begin{align}\label{det}
\begin{aligned}
  |\det R_i|& = |\mu_{i11}^{m-1}\mu_{i21}^{m-2}-\mu_{i11}^{m-2}\mu_{i21}^{m-1}|^2
|\det V_0|^2|\det R_0|\v^{m^2-3m+3}(1+o(1)) \\
  & = |\mu_{i11}^{m-2}\mu_{i21}^{m-2}(\mu_{i11}-\mu_{i21})|^2
|\det V_0|^22^m\v^{m^2-3m+3}(1+o(1)),
\end{aligned}
\end{align}
where $V_0=V(\vartheta_{31},\dots,\vartheta_{m1})=\left[
             \begin{array}{cccc}
               1 & \vartheta_{31} & \dots & \vartheta_{31}^{m-3} \\
               1 & \vartheta_{41} & \dots & \vartheta_{41}^{m-3} \\
               \vdots & \vdots & \ddots & \vdots \\
               1 & \vartheta_{m1} & \dots & \vartheta_{m1}^{m-3} \\
             \end{array}
           \right]
$ is a Vandermonde matrix of order $m-2$.

On the other hand, we need to determinate the order of the cofactor $C^{(i)}_{s,t}$
of the $(s,t)$ entry, and we illustrate it by calculating $C^{(i)}_{1,1}$ with $m=3$.
After deleting the first column $u_{i1}$, we delete the first row
and use the same notations $u_1,u_2$ and $e_2,\dots,e_6$.
Now $R_i$ has been reduced to a square matrix $R^{(i)}_{1,1}$ consisting of the following
5 columns:
$$
\begin{array}{l}
  a_1=-j e_2+\v^{1/2}\bar\mu_{i11}\bar u_1+\v\bar\mu_{i11}^2\bar u_2+O(\v)\bar{u}_1+O(\v^2),\\
  a_2= j e_2+\v^{1/2}\mu_{i21} u_1+\v\mu_{i21}^2 u_2+O(\v){u}_1+O(\v^2),~a_3=\bar a_2,\\
  a_4= j e_2+\v\vartheta_{31} u_1+\v^2\vartheta_{31}^2 u_2+O(\v^2){u}_1+O(\v^2),~a_5=\bar a_4. \\
\end{array}
$$
Consequently the order of $C^{(i)}_{1,1}$ is found in such a way:
take out terms with $\v^{1/2}\bar u_1,\v\bar u_2$ respectively from $a_1,a_3$,
terms with $e_2$ from $a_5$, terms with $\v u_2$ from $a_2$, terms with $\v u_1$ from $a_4$.
Now that $a_1,a_3,a_5$ jointly contribute the same degree of $\frac{1}{2}(m^2-3m+3)$ as
$\bar u_{i1},\bar u_{i2},\bar u_{i3}$,
we are left to choose terms with $u_1$ and $u_2$ respectively from $a_2$ and $a_4$.
The above can be conducted similarly for calculating the order of $C^{(i)}_{1,1}$ when $m>3$,
and actually for every cofactor.
Moreover, by the symmetry of conjugates,
$C^{(i)}_{2k-1,2n-1},C^{(i)}_{2k-1,2n},C^{(i)}_{2k,2n-1},C^{(i)}_{2k,2n}$
have an identical order. So we only focus on $C^{(i)}_{2k-1,2n-1}$ below.
Reminded by the case of $C^{(i)}_{1,1}$ when $m=3$,
we suffice to choose linearly independent terms with a lowest sum of degrees from
the modified columns $u_{ij}$ for $j\ne n$, where $u_{k-1}$ has been subtracted from each column.
Recall that in finding the order of $|\det R_i|$,
terms with $\v u_0,\dots,\v^{m-3}u_{m-3}$ from type II columns are first selected,
and then terms with $\v^{(m-2)/2}u_{m-2},\v^{(m-1)/2}u_{m-1}$ from type I columns.
Such a method still applies in finding the order of cofactors,
and we conclude that $C^{(i)}_{2k-1,2n-1}$ has the lowest order for fixed $n$
if and only if $k=m$.
In other words, for any row in adj$R_i$, the entries at the $2m-1$-th and $2m$-th column
exclusively have the lowest order when compared with other entries at the same row.
To be detailed,
\begin{equation}\label{cofacnor}
\begin{array}{l}
\quad|C^{(i)}_{2m-1,2n-1}|,|C^{(i)}_{2m-1,2n}|,|C^{(i)}_{2m,2n-1}|,|C^{(i)}_{2m,2n}|\\
=\left\{
   \begin{array}{ll}
     2^{m-1}|\det V_0|^2|\mu_{i11}^{m-2}\mu_{i21}^{m-2}(\mu_{i11}-\mu_{i21})|
|\mu_{in1}^{m-2}|\\
\cdot\v^{m^2-3m+3-(m-1)/2}(1+o(1)), ~n=1,2; \\
     2^{m-1}|\det V_0||\det V_n||\mu_{i11}^{m-2}\mu_{i21}^{m-2}(\mu_{i11}-\mu_{i21})|\\
\cdot|\mu_{i11}^{m-3}\mu_{i21}^{m-3}(\mu_{i11}-\mu_{i21})|\v^{m^2-3m+3-(m-2)}(1+o(1)), \; n=3,\dots,m,
   \end{array}
 \right.
\end{array}
\end{equation}
where $V_n=V(\vartheta_{31},\dots,\vartheta_{n-1,1},\vartheta_{n1},\dots,\vartheta_{m1})$ is a Vandermonde matrix of order $m-3$.
Together with \eqref{det} it yields that by $\mu_{i11}=-\mu_{i21}$
\begin{equation}\label{invRi}
\begin{array}{l}
\quad|R^{-1}_{i}(2n-1,2m-1)|,|R^{-1}_{i}(2n,2m-1)|,|R^{-1}_{i}(2n-1,2m)|,|R^{-1}_{i}(2n,2m)|\\
=\left\{
   \begin{array}{ll}
\frac{1}{4|\mu_{i11}|^{m-1}}\v^{-(m-1)/2}(1+o(1)), ~n=1,2; \\
     \frac{1}{2|\mu_{i11}|^{2}}\frac{|\det V_n|}{|\det V_0|}\v^{-(m-1)}(1+o(1)), \; n=3,\dots,m;
   \end{array}
 \right.\\
\quad|R^{-1}_{i}(2n-1,k)|,|R^{-1}_{i}(2n,k)|\\
=\left\{
   \begin{array}{ll}
o(\v^{-(m-1)/2}), ~n=1,2; \\
o(\v^{-(m-1)}), \; n=3,\dots,m;
   \end{array}
 \right.
\text{for }1\le k\le 2m.
\end{array}
\end{equation}

In the meanwhile, the following holds for $1\le j\le m$:
\begin{equation}\label{Ri}
\begin{array}{l}
\quad|R_{i}(2j-1,2n-1)|,|R_{i}(2j-1,2n)|,|R_{i}(2j,2n-1)|,|R_{i}(2j,2n)|\\
=\left\{
   \begin{array}{ll}
|\mu_{in1}^{j-1}|\v^{(j-1)/2}+O(\v^{j/2}), ~n=1,2; \\
     |\vartheta_{n1}^{j-1}|\v^{j-1}+O(\v^{j}), \; n=3,\dots,m;
   \end{array}
 \right.\\
\end{array}
\end{equation}

Combining (\ref{Aij}), (\ref{invRi}) and (\ref{Ri}) we can obtain
\begin{align*}
&\quad |\xi_{s,2j-1}|\\
&\le \rho_{i}^s||\xi||\big[
\sum_{n=1}^{m}|R_i(2j-1,2n-1)|(|R^{-1}_{i}(2n-1,2m-1)|+|R^{-1}_{i}(2n-1,2m)|)\\
&\quad +\sum_{n=1}^{m}|R_i(2j-1,2n)|(|R^{-1}_{i}(2n,2m-1)|+|R^{-1}_{i}(2n,2m)|)
\big](1+o(1))\\
& \le \rho_{i}^s||\xi||2\big[
\sum_{n=1}^{2}\frac{|\mu_{in1}^{j-1}|}{4|\mu_{i11}^{m-1}|}\v^{(j-m)/2}
+\sum_{n=3}^{m}\frac{|\vartheta_{n1}^{j-1}|}{2|\mu_{i11}^{2}|}\frac{|\det V_n|}{|\det V_0|}
\v^{j-(m-1)}\big](1+o(1)).
\end{align*}
and the conclusion follows by noticing
that $j-(m-1)<\frac{j-m}{2}$ for $j<m-2$, $j-(m-1)=\frac{j-m}{2}$ for $j=m-2$
and $j-(m-1)>\frac{j-m}{2}$ for $j=m-1,m$,
as well as $|\mu_{i11}|=\sqrt{\lambda_i/2}$,
$\frac{|\det V_n|}{|\det V_0|}=\prod\limits_{3\leq k\leq m,k\neq n}|\vartheta_{k1}-\vartheta_{n1}|$.   \IEEEQEDclosed

{\em Proof of Lemma \ref{lem3}}
The proof of Lemma \ref{lem3} relies on the following combinatorial identity.
\begin{lem}\label{comiden}\cite{sprugnoli_riordan_2007}
Let $f(t)=\sum_{k=0}^{\infty}f_kt^k$ be a formal power series \cite{sprugnoli_introduction_2006}.
Then the following rule holds if $b=0$ and $f(t)$ is a polynomial:
$$
\sum_{k}C(n+ak,m+bk)z^{m+bk}f_k=[t^m](1+zt)^nf(t^{-b}(1+zt)^a),
$$
where $[t^m]g(t)$ denotes the extraction of the coefficient of $t^m$ from the formal power series $g(t)$.
\end{lem}
Now let we return to the proof.
Denoting
\begin{align*}
  v(\t)& = \cos\t A^{2m-1}(1,\cdot)+\sin\t A^{2m-1}(2,\cdot)+A^{2m-1}(3,\cdot) \\
  & = [v_{1}(\t)~v_{2}(\t)\dots v_{2m-1}(\t)~v_{2m}(\t)]
\end{align*}
and recalling $S=A^{2m-1}\mathcal {O}^{-1}$, the original equation is equivalent to
$l(\t)\mathcal {O}=v(\t)$.
Direct computation shows that the entries of $v(\t)$ are given by
\begin{align*}
  v_{2j-1}(\t)&=C(2m,j-1)\cos(2m-j+1)\t, \\
  v_{2j}(\t)&=C(2m,j-1)\sin(2m-j+1)\t,~j=1,\dots,m;
\end{align*}
and the entries of $\mathcal {O}$ are given by
\begin{align*}
  \mathcal {O}(k,2j-1)&=C(k-1,j-1)\cos (k-j)\t, \\
  \mathcal {O}(k,2j)&=C(k-1,j-1)\sin (k-j)\t,~k=1,\dots,2m,~j=1,\dots,m.
\end{align*}
As a result, the equation $l(\t)\mathcal {O}=v(\t)$ is equivalent to the following $m$ equations:
\begin{equation}\label{eqorigin}
\sum_{k=0}^{2m-1}l_k(\t)C(k,h)e^{\jmath(k-h)\t}=C(2m,h)e^{\jmath(2m-h)\t},
~h=0,1,\dots,m-1,
\end{equation}
or equally
\begin{equation}\tag{\ref{eqorigin}$'$} \label{eqprime}
\sum_{k=0}^{2m}l_k(\t)C(k,h)e^{\jmath(k-h)\t}=0,~h=0,1,\dots,m-1,
\end{equation}
if we let $l_{2m}(\t)=-1$.
Noticing that $2\cos\t=e^{\jmath\t}+e^{-\jmath\t}$,
we substitute the expression of $l_k(\t)$ into the left-hand side of the above $h$-th equation,
and expand it into a power series of $e^{\jmath\t}$ as
$\sum_{k=0}^{2m}l_k(\t)C(k,h)e^{\jmath(k-h)\t}=\sum_{w=0}^{m}\alpha_{w,h}e^{\jmath(2w-h)\t}$,
with
$$
\alpha_{w,h}=\sum_{k=0}^{w}\sum_{j=w-k}^{m-k}C(m,j)C(m-j,k)C(j,w-k)C(j+2k,h)(-1)^{j+2k-1}.
$$
Therefore, if we can show that $\alpha_{w,h}$ for $w=0,1,\dots,m$ and $h=0,1,\dots,m-1$,
then the prescribed $l(\t)$ is a solution of \eqref{eqorigin},
and by the nonsigularity of $\mathcal{O}$ it is also unique.

We first transform $\alpha_{w,h}$ as follows.
By remembering that
\begin{align*}
C(m,j)C(m-j,k)C(j,w-k)&=C(m,k)C(m-k,j)C(j,w-k)\\
&=C(m,k)C(m-k,w-k)C(m-w,j-(w-k))
\end{align*}
and letting $s=j-(w-k)$, it is clear that
$$
\alpha_{w,h}=\sum_{k=0}^{w}C(m,k)C(m-k,w-k)(-1)^{k+w-1}
\sum_{s=0}^{m-w}C(m-w,s)C(s+w+k,h)(-1)^s.
$$
Now we claim that

\noindent1). $\sum_{s=0}^{m-w}C(m-w,s)C(s+w+k,h)(-1)^s=C(w+k,h-(m-w))(-1)^{m-w}$, $w=0,1,\dots,m,$

\noindent2). $\sum_{k=0}^{w}C(m,k)C(m-k,w-k)C(w+k,h-(m-w))(-1)^{m+k-1}=0$, $h=0,1,\dots,m-1$,

\noindent and the proof of the first part is completed by combining these two claims.

1). Let $f_s=(-1)^sC(m-w,s)$ and $f(t)=\sum\limits_{s}f_st^s=(1-t)^{m-w}$.
Applying Lemma \ref{comiden}, we have
\begin{align*}
    &\quad \sum_{s=0}^{m-w}C(m-w,s)C(s+w+k,h)(-1)^s \\
  & = \sum_{s=0}^{m-w}C(w+k+1\cdot s,h+0\cdot s)\cdot 1\cdot f_s\\
  & = [t^h](1+t)^{w+k}f(1+t)\\
  & = [t^h](1+t)^{w+k}(-t)^{m-w}\\
  & = (-1)^{m-w}[t^{h-({m-w})}](1+t)^{w+k}\\
  & = (-1)^{m-w}C(w+k,h-(m-w)),
\end{align*}
which establishes the first claim.

2).
For the second claim,
\begin{align*}
    & \quad \sum_{k=0}^{w}C(m,k)C(m-k,w-k)C(w+k,h-(m-w))(-1)^{m+k-1} \\
  & = (-1)^{m-1}\sum_{k=0}^{m}C(m,k)(-1)^{k}C(m-k,w-k)C(w+k,h-(m-w)) \\
  & = (-1)^{m-1}\sum_{k=0}^{m}[t^k](1-t)^m[v^{m-w}](1+v)^{m-k}[u^{h-(m-w)}](1+u)^{w+k}\\
  & = (-1)^{m-1}[v^{m-w}](1+v)^{m}[u^{h-(m-w)}](1+u)^{w}\sum_{k=0}^{m}[t^k](\frac{1+u}{1+v})^k(1-t)^m\\
  & = (-1)^{m-1}[v^{m-w}](1+v)^{m}[u^{h-(m-w)}](1+u)^{w}(1-\frac{1+u}{1+v})^m\\
  & = (-1)^{m-1}[v^{m-w}](1+v)^{m}[u^{h-(m-w)}](1+u)^{w}\frac{(v-u)^m}{(1+v)^m}\\
  & = (-1)^{m-1}[v^{m-w}][u^{h-(m-w)}](1+u)^{w}(v-u)^m\\
  &\xlongequal{r=h-(m-w)}  (-1)^{m-1}[v^{m-w}][u^{r}](1+u)^{w}(v-u)^m\\
  & = (-1)^{m-1}[v^{m-w}]\sum_{k=0}^{r}C(w,r-k)C(m,k)v^{m-k}(-1)^k\\
  & = (-1)^{m-1}[v^{k-w}]\sum_{k=0}^{r}C(w,r-k)C(m,k)(-1)^k,
\end{align*}
where $t,v,u$ are indeterminates.
Noticing that $r\geq w\Leftrightarrow h-(m-w)\geq w\Leftrightarrow h\geq m$ is
contradictory to $h=0,1,\dots,m-1$,
we have $r<w$ and $k\le r<w$, which suggests the vanishing of the last equation in the above,
and the proof for the first part is complete.

As for the second part, by noting that the exponents of $\cos\t$ in $l_k(\t)$ are even
when $k$ is an even number,
while the exponents are odd when $k$ is an odd number,
it can be noted that the sign of each term in $l_k(\t)$ is the same.
Therefore we obtain
\begin{align*}
  \sum_{k=0}^{2m-1}|l_k(\t)|& =
 \sum_{k=0}^{2m-1}\sum_{h=0}^{\lfloor k/2\rfloor }C(m,k-2h)C(m-(k-2h),h)|2\cos\t|^{k-2h} \\
  & = \sum_{j=0}^m\beta_j |\cos\t|^j,
\end{align*}
with $\beta_0=2^m-1$ and $\beta_j=2^jC(m,j)\sum_{h=0}^{m-j}C(m-j,h)=2^mC(m,j)$ for $j=1,\dots,m$,
and the conclusion follows directly. \IEEEQEDclosed

{\bf Acknowledgement}
The authors would like to thank Dr. Shuai Liu for his valuable suggestions.

\bibliographystyle{IEEEtran}
\bibliography{oscillator_arXiv}

\end{document}